\documentclass[10pt]{article}
\usepackage{amssymb}
\usepackage{amstext}
\newcommand*\de{\mathrm{d}}
\newcommand*\De{\mathrm{D}}
\renewcommand*\epsilon{\varepsilon}
\renewcommand*\phi{\varphi}
\renewcommand*\theta{\vartheta}

\begin{document}
  
\title{\bf  Canonical and gravitational stress-energy tensors} 
\author{\small M. Leclerc\\\small  
Section of Astrophysics and Astronomy, 
Department of Physics, \\ \small University of Athens, Greece}  
\date{\small April 11, 2006}
\maketitle
\begin{abstract}
It is dealt with the question, under which circumstances the 
canonical Noether stress-energy tensor is equivalent to the 
gravitational (Hilbert) tensor for general matter fields under 
the influence of gravity. In the framework of general relativity, 
the full equivalence is established for  matter fields 
that do not couple to the metric derivatives.  
Spinor fields are included into our analysis by reformulating 
general relativity in terms of tetrad fields, and the case of 
Poincar\'e gauge theory, with an additional, independent Lorentz 
connection, is also investigated. Special attention is given to 
the flat limit, focusing  on the expressions for the 
matter field  energy  (Hamiltonian).  
The Dirac-Maxwell 
system is investigated in detail, with special care given to the 
separation of free  (kinetic) and interaction (or potential) energy. 
Moreover, the 
stress-energy tensor of the gravitational field itself is briefly discussed.  

\end{abstract}

\section{Introduction}
In textbooks on general relativity, the Hilbert stress-energy tensor 
is often presented as an improvement over the canonical Noether 
tensor,  because it is automatically symmetric, while the Noether 
tensor has to be submitted to a relocalization if one insists on 
a symmetric tensor. That we have, on one hand, a symmetric tensor, 
and on the other hand, a tensor that is physically equivalent to a symmetric
tensor, is thus well known. This, however, does still not proof that 
both tensors are indeed equivalent. Especially, it remains unclear if 
 the symmetrized 
Noether tensor is generally identical to the Hilbert tensor.  
Unfortunately, in 
literature, the only cases which are explicitely treated 
are the scalar and the free electromagnetic field. 
 Only recently, attention has been paid to the 
general relationship between different concepts of 
stress-energy\cite{1,2,3,4}, most authors concluding on  equivalence. However, 
this is obviously not always the case (see Maxwell field with 
sources), and therefore, we see the need of a complete investigation of 
the subject.    

We carry out an analysis of the relations between 
both conceptions of stress-energy without  reference to the specific  
nature of the matter fields involved. We will perform this analysis in 
the framework of general relativity (section 2), Poincar\'e gauge 
theory (section 4)  
and tetrad gravity (section 5). The canonical stress-energy of both 
gravitational and matter fields  is briefly discussed in section 3.  
Finally, we treat the specific examples of a classical point charge in 
general relativity (section 6) and the Dirac-Maxwell system, 
where we focus on the flat limit expression of the field energy 
(Hamiltonian), giving 
special attention to the separation of free and interaction part, 
as well as to the 
case where part of the fields are considered to be 
background fields, as is most common in practical applications (section 7). 

We should warn the reader that the aim of this article is not 
to produce any kind of new, spectacular results, but rather to 
clarify the old concepts of canonical and gravitational approaches 
and to point out  potential problems that  eventually can 
lead to misconceptions.

\section{Classical general relativity}

Let the complete action be composed of the gravitational part $- 1/2 \int 
\hat R \sqrt{-g}\ \de^4 x$ plus the matter part 
\begin{equation} \label{1}
S = \int \sqrt{-g} L\ \de^4 x, 
\end{equation}
with $L = L(q)$ depending on the matter fields denoted collectively 
by $q = (A_i, \psi, \bar \psi, \phi \dots)$ and their first derivatives, 
as well as on the metric tensor $g_{ik}$. Our  conventions 
correspond to the standard Landau-Lifshitz\cite{5} notations, 
except for the hat 
on the Christoffel symbols $\hat \Gamma^i_{lm}$ and the curvature tensor 
formed from them, $\hat R^i_{\ klm}$. The gravitational field equations are 
easily derived by variation with respect to $g_{ik}$ and read 
$\hat G^{ik} = T^{ik}$, where the gravitational (Hilbert, or metric) 
stress-energy tensor is defined as 
\begin{eqnarray}\label{2}
T^{ik} &=& -2 \frac{1}{\sqrt{-g}} \frac{\delta (L \sqrt{-g})}{\delta g_{ik}}
\nonumber \\
&=& -2 \frac{1}{\sqrt{-g}} \frac{\partial (L \sqrt{-g})}{\partial  g_{ik}}
+  2 \frac{1}{\sqrt{-g}} \partial_{m} 
(\frac{\partial (L \sqrt{-g})}{\partial g_{ik,m}}). 
\end{eqnarray}
In most 
  applications, the second term will be absent, because the matter
fields usually couple only to the metric and not to its derivatives. This 
is the case for the Maxwell field and gauge fields in general, as well as 
for scalar fields. Spinor fields need a different treatment and are excluded 
in this section. We therefore omit the second term for the moment. Let us 
also note that $T^{ik}$ is, by construction, symmetric.

An immediate consequence of the field equations, using the Bianchi identities 
of the curvature tensor, is the covariant conservation law 
$T^{ik}_{\ \ ;k} = 0$. The fact that this is not a conservation law in the 
strict sense is explained in any textbook on general relativity. For 
convenience, however, we will call \textit{covariant conservation law} any
covariant relation that reduces, in the flat limit, to a conservation law. 
More instructive than using the field equations is the derivation of this 
relation based on the invariance under coordinate transformations 
of the matter action. Especially, it makes clear exactly in which cases  
 the relation really holds and in which it does not. 

Consider an infinitesimal coordinate transformation 
$x^i \rightarrow \tilde x^i = x^i + \xi^i$. 
Then, the invariance condition reads
\begin{equation}\label{3}
\delta S = \int (\frac{\delta (\sqrt{-g} L)}{\delta q} \delta q 
+ \frac{\delta (\sqrt{-g} L)}{\delta g_{ik}} \delta g_{ik}) \de^4 x = 0. 
\end{equation}
 The crucial point is that, by means of the matter field equations, 
we have $\delta (\sqrt{-g}L)/ \delta q = 0 $, and therefore the first 
term vanishes. It is important to stress that this holds only if we 
consider the complete matter action. If, for instance,  we consider 
a system of a test particle in a central electric field,  
 we usually consider 
the Lagrangian containing the particle's field and the interaction term of 
the particle with the Maxwell field, but we omit the free field 
part $\sim F^2$. In such a case, it cannot be expected that the gravitational 
stress-energy tensor be covariantly conserved (because not all field 
equations are exploited). We will illustrate this for the concrete case  of 
the Dirac-Maxwell system later on. 

As to the second term, we use the transformation law 
\begin{equation} \label{4}
\delta g_{ik} = \tilde g_{ik}(x) - g_{ik}(x) = - \xi_{i;k} - \xi_{k;i}, 
\end{equation}
which is the result of $g_{ik}$ transforming as a tensor, and expressing 
the transformed metric in terms of the old coordinates (see Ref. \cite{5}). 
Inserting this into  (\ref{3}), using the definition (\ref{2}) and partially
integrating by omitting a surface term, leads, in view of the independence of 
the transformation coefficients $\xi^i$, to 
\begin{equation}\label{5}
T^{ik}_{\ ;k} = 0. 
\end{equation}
 
On the other hand, in the context of special relativistic field theory, 
the following, canonical, or Noether, stress-energy tensor is widely in use: 
\begin{equation} \label{6}
\tau^k_{\ i} 
= \frac{\partial L}{\partial q_{,k}} q_{,i} -
  \delta^k_i  L, 
\end{equation}
and it is then shown that, as a result of the matter field equations, 
this tensor is conserved, i.e., $ \tau^k_{\ i,k}= 0$. This holds 
in flat spacetime. In general, $\tau^{ki}$ is not symmetric. In a next
step, most textbooks on field theory  
show that the canonical tensor can be rendered symmetric 
by a so-called relocalization, which does not affect the conservation law 
$\tau^k_{\ i,k} = 0$ nor the canonical momentum vector defined by   
\begin{equation}\label{7}
\pi_i = \int \sqrt{-g}\  \tau^0_{\ i}\ \de^3 x. 
\end{equation}
(The trivial factor $\sqrt{-g}$  is only introduced for later considerations.) 
It is unnecessary to recall that, as a result of  $\tau^k_{\ i,k} = 0$, 
we also have $\de P_i / \de t = 0$, which is just the integral form 
of the conservation law. The point is that such relocalizations do not 
affect the  quantities $P_i$ and are therefore irrelevant from  a 
physical point of view. 

Let us also mention  that, in contrast to what is often stated in 
literature (for instance in Ref. \cite{5}), 
the relocalization procedure is not unique, even 
under the requirement that the final stress-energy tensor  
be symmetric. For instance, to the canonical 
 tensor $\tau^{ik}$ of a scalar field 
in flat spacetime, which is already symmetric, 
we can add the relocalization term  $[\phi^{,k} \phi\ \eta^{il} - 
\phi^{,l}\phi\  \eta^{ik}]_{,l}$, which is again symmetric and has an  
identically vanishing divergence. There exists, however, 
 a more consistent procedure, the so-called Belinfante 
symmetrization, that leads to a uniquely defined symmetric 
stress-energy tensor (based on the canonical tensor), and 
moreover, has the nice property that it is insensitive to 
surface terms in the action, which is not the case, e.g., for 
the tensor (\ref{6}). There is, however, a price to pay. The Belinfante 
procedure relies on the Noether current corresponding  to 
global Poincar\'e (coordinate) transformations. Certainly, 
any diffeomorphism invariant action will also be globally 
Poincar\'e invariant, but there is no apparent  need, a priori, to  
favor a certain subgroup. In our opinion, this is against the spirit 
of general relativity. (For instance, in general relativity with 
cosmological constant, the de Sitter subgroup is  at least 
equally well justified.) 
For an  illuminating discussion on this procedure, 
we refer to Ref. \cite{jackiw}. 

Let us also emphasize  that, whenever we mention the canonical stress-energy 
tensor, we refer explicitely to the expression \ref{6}. This is simply 
 because 
it is directly related to the canonical formalism in quantum field theory 
(recall, e.g.,  the so-called canonical momentum $\partial  L / \partial 
q_{,m}$ as well as the canonical construction of the
Hamiltonian).  We caution that in literature, there are also other expressions 
 that are refereed to as \textit{canonical} stress-energy tensor, because 
they also arise from a canonical procedure (albeit a different one). 
In particular, in Ref. \cite{mont1}, an interesting procedure 
has been described where the symmetric and gauge invariant 
 expressions for the stress-energy tensor of the Maxwell and Yang-Mills 
fields can be obtained directly from Noethers identities, taking into 
account, apart from the field equations, also the Bianchi equations 
related to gauge invariance. It would be of interest, especially in 
view of Poincar\'e gauge theory (see section 4) 
 to carry out a similar analysis for 
the case of the Dirac field, which allows for a local  Lorentz gauge
invariance. Another example  is the 
Hilbert stress-energy tensor (\ref{2}), to which we refer here 
as gravitational (or metric), and which is also often referred to as 
\textit{canonical}.

In most textbooks  on general relativity, it is shown that the metrical tensor 
coincides, up to a relocalization, with the canonical tensor as far as 
 scalar fields and the electromagnetic fields are concerned (although, 
in the later case, this is true only for the sourceless fields). 
What cannot be found in textbooks is a general proof  that the 
metrical tensor (when considered in the flat limit) is always  equivalent
with (i.e., equal to,  up to a relocalization)  the canonical stress-energy 
tensor, without referring to specific matter fields. 

What is also unclear is the generalization from flat to curved spacetime. 
If we define the canonical stress-energy tensor as before, namely by 
(\ref{6}), 
it cannot be expected that we still have $\tau^k_{\ i,k}= 0$, but neither 
$\tau^k_{\ i;k}$, since, in general, $\tau^k_{\ i} $ is not even a covariant 
object (nevertheless, we will continue to refer to it as canonical 
\textit{tensor}) 
and therefore the use of a covariant derivative does not seem to make 
much sense. 

In the following, we will try to clarify those points. First, recall that 
from (\ref{1}), we derive the matter field equations in the form 
\begin{equation}\label{8}
\frac{\partial}{\partial q} ( \sqrt{-g} L) 
= \partial_i (\frac{\partial(\sqrt{-g}L)}{\partial q_{,i}}).
\end{equation}
Next, we evaluate the partial derivative of the Lagrangian density 
\begin{equation}\label{9}
\frac{\partial}{\partial x^i} (L \sqrt{-g}) = 
\frac{\partial(L\sqrt{-g})}{\partial q} q_{,i} 
+ \frac{\partial(L\sqrt{-g})}{\partial q_{,k}} q_{,k,i} 
+ \frac{\partial(L\sqrt{-g})}{\partial g_{kl}} g_{kl,i}.  
\end{equation}
Recall that, for the moment, we consider $L$ not to depend on the metric 
derivatives. In the first term of the r.h.s., we use the field equations 
and find 
\begin{equation}\label{10}
0 = \partial_k \left[\frac{\partial(L\sqrt{-g})}{\partial q_{,k}} q_{,i}-
\delta^k_i (L \sqrt{-g} )\right] + \frac{\partial(L\sqrt{-g})}{\partial g_{lm}}
g_{lm,i}. 
\end{equation}
Using the definitions (\ref{2}) and (\ref{6}), we can write 
\begin{equation}\label{11}
0 = \partial_k[ \sqrt{-g}\  \tau^k_{\ i}] - \frac{1}{2} \sqrt{-g}\  
T^{lm} g_{lm,i}.  
\end{equation}
Clearly, this has to be considered to be the generalization of the 
conservation law $\tau^k_{\ i,k}=0$ of the canonical stress-energy tensor, 
to which it reduces immediately for $g_{lm,i} = 0$. It does not look very 
useful, since two stress-energy tensors are involved. Let us 
rewrite (\ref{11}) 
in the form 
\begin{equation} \label{12}
0 = \partial_k \left[ \sqrt{-g}\  (\tau^k_{\ i}- T^k_{\ i})\right]
+ \partial_k (\sqrt{-g} T^k_{\ i}) 
- \frac{1}{2} \sqrt{-g}\  T^{lm} g_{lm,i}.  
\end{equation}
It is not hard to show that the last two terms are just $\sqrt{-g}\ T^k_{\
  i;k}$, which vanishes as we have already shown. Our final relation 
therefore reads 
\begin{equation}\label{13}
0 = \partial_k \left[ \sqrt{-g}\  (\tau^k_{\ i}- T^k_{\ i})\right]. 
\end{equation}
Note that this is not a conservation law (i.e., something related to 
equations of motion), but a relation that is identically fulfilled in 
view of the conservation laws (\ref{11}) and (\ref{5}). 
It simply determines the 
relation between the canonical and the metrical stress-energy tensor. It is 
exactly what we need in order 
to show the equivalence of both tensors, since from (\ref{13}), 
it follows for the canonical  momentum (\ref{7}) and 
the corresponding gravitational momentum 
\begin{equation}\label{14}
P_i = \int \sqrt{-g}\ T^0_{\ i}\ \de^3 x, 
\end{equation}
that we have 
\begin{equation} \label{15}
\frac{\de }{\de t} (\pi_i - P_i) = 0, 
\end{equation}
i.e., $P_i$ and $\pi_i $ coincide up to an irrelevant constant. 
We should 
caution that  we use the term \textit{gravitational} stress-energy 
tensor and  \textit{gravitational} momentum for the quantities that arise 
upon variation (of the matter Lagrangian) 
with respect to the gravitational field $g_{ik}$ (or 
$e^a_i$ in the next sections). This is not to be confused with 
the stress-energy tensor or the momentum of the gravitational field 
itself. Unless otherwise stated, we always refer to the matter contributions.
With (\ref{15}), we 
have shown in full generality, that in curved spacetime, the canonical 
and the metric stress-energy tensors are physically equivalent. Nevertheless, 
one should not forget that in general, $\tau^k_{\ i}$ is not a covariant 
object (consider the Maxwell case for instance), 
and the relation $\tau^k_{\ i;k}= 0$ only holds in special cases, 
where $T^{ik} = \tau^{ik}$, as is the case, e.g., for the scalar field.  
And to avoid any misunderstanding, by physically equivalent, we mean of 
course equivalent, as far as the evolution of the matter fields is 
concerned. Only $T^{ik}$ can be used as source term for the gravitational 
field equations.

As to the constant of integration arising in (\ref{15}), 
although  irrelevant on a classical level,  it is interesting that 
it vanishes anyway. Indeed, as outlined above, equation (\ref{13})
is actually an identity. This is simply because  we have 
already used all the field equations, and thus, no farther constraints 
can arise on our fields. This means that $\sqrt{-g}(\tau^k_{\ i} - T^k_{\ i})$
has to be a relocalization term, i.e., it is of the form 
$r^{kl}_{\ \ i ,l}$ for some object $r^{kl}_{\ \ i}$ satisfying 
$r^{kl}_{\ \ i}= -r^{lk}_{\ \ i}$. 
It is not hard to show that for such an object, 
$\int r^{0l}_{\ \ i,l}\ \de^3 x $ vanishes identically (up to 
surface terms). Therefore, we conclude that  
\begin{equation} \label{16}
\pi_i = P_i.
\end{equation}
This completes our argument.

Although standard model fields do not couple to the metric derivatives (for 
spinors, see sections 4 and 5), in 
the framework of general relativity, nothing 
really prevents us from considering, e.g., a massive vector field  
whose  kinetic Lagrangian is of the form $\sim B_{i;k} B^{i;k}$ or 
similar. In such cases, we have to add in equation (\ref{9}) the term 
$(\partial (\sqrt{-g}L)/ \partial g_{lm,k}) g_{lm,k,i} $ and, when 
performing the step 
 from 
(\ref{10}) to (\ref{11}), include the second term from (\ref{2}) 
in the definition of $T^{ik}$. 
The relation (\ref{5}) 
still holds, it has been derived without restrictions on 
$L$. As a result, instead of (\ref{13}), we find 
\begin{equation}\label{17}
0 = \partial_k \left[ \sqrt{-g}\  (\tau^k_{\ i}- T^k_{\ i} + \frac{\partial
  L}{\partial g_{lm,k}} g_{lm,i})\right],    
\end{equation}
which means that in this case, apart from a  constant that 
can be shown to be zero again, 
\begin{equation}\label{18}
\pi_i = P_i - \int  \frac{\partial (\sqrt{-g} L)}{\partial g_{lm,0}} g_{lm,i}
\ \de^3 x,  
\end{equation}
which  shows the non-equivalence of 
$T^{ik}$ and $\tau^{ik}$, which is restored, however, in the flat limit. 
(Be careful: In some papers, especially in  Refs. \cite{3,4}, 
the term \textit{canonical} tensor is 
used, not for the expression (\ref{6}), but rather for a similar expression, 
but with the partial derivatives replaced with a covariant derivative. 
Therefore, the result of Ref. \cite{3} is not in contradiction with ours, 
but simply refers to a completely different tensor.)  

Although the case where  the 
metric derivatives couple to matter 
is rather unlikely to occur in classical general 
relativity, 
a similar situation occurs when dealing with spinor fields, where the 
metric is replaced with a tetrad field. 

Before, however, we would like to shortly discuss the stress-energy 
tensor of the gravitational field itself, which is often treated with 
some obscurity.

\section{Canonical stress-energy tensor for gravity}

As mentioned above, the covariant conservation law $T^{ik}_{\ ;k} $ 
of the metric stress-energy tensor of the matter field is not 
a conservation law in the sense that it leads to a conserved momentum 
vector. The reason can be seen in the fact that only the sum 
of the matter momentum and the momentum of the gravitational field 
itself is conserved (see Ref. \cite{5}, \S 96). (Alternatively, one could 
say that the momentum is not conserved because it is subject to 
gravitational forces.) In Ref. \cite{5}, it is shown how one can derive 
a quantity $t^{ik}$, depending only on the metric (and its derivatives) 
such that the following momentum (pseudo) vector 
\begin{equation}\label{19}
\Pi^i = \int (-g) (T^{i0} + t^{i0})\de^3 x 
\end{equation}
is conserved.  The quantity $t^{ik}$ is known as the Landau-Lifshitz 
pseudo-stress-energy tensor for the gravitational field. The derivation of 
(\ref{19}) 
is rather involved. The reason is that the authors took care that the 
resulting pseudo-tensor $t^{ik}$ be symmetric, in order for the 
angular-momentum tensor to be conserved. In our viewpoint, the symmetry 
of the stress-energy tensor is not of too much importance, especially 
as it will be lost anyway when we go over to Poincar\'e gauge theory 
in the next section and we would rather prefer to have conservation laws 
for a momentum that includes either (\ref{7}) or (\ref{14}) 
as far as the matter part of 
the momentum is concerned. This is not the case with (\ref{19}), 
because of the 
factor $(-g)$ instead of $\sqrt{-g}$. 

There is an easy solution to this, which the authors of Ref. 
\cite{5} choose to 
hide in the  problem section of \S 96. We will briefly resketch it  here 
in order to show the complete similarity with the derivations made earlier, 
going from (\ref{9}) to (\ref{11}). 

Let $L_{tot} = L_{grav} + L$, where $L$ is, as before, the matter Lagrangian, 
and $L_{grav}$ the free gravitational part. Consider the derivative of 
$L_{tot}$, 
\begin{eqnarray}\label{20}
(\sqrt{-g} L_{tot})_{,i} &=& \frac{\partial (\sqrt{-g}L_{tot})}
{\partial q} q_{,i}
+\frac{\partial (\sqrt{-g}L_{tot})}{\partial q_{,m}} q_{,m,i}
\nonumber \\&&
+\frac{\partial (\sqrt{-g}L_{tot})}{\partial g_{lm}} g_{lm,i}
+\frac{\partial (\sqrt{-g}L_{tot})}{\partial g_{lm,k}} g_{lm,k,i}, 
\end{eqnarray}
Apart from the matter field equations (\ref{8}), which can be used in the 
first term, we now also  have at our disposal the gravitational field 
equations 
in the form 
\begin{equation}\label{21}
\frac{\partial}{\partial g_{lm}} (\sqrt{-g} L_{tot}) 
= \partial_i (\frac{\partial(\sqrt{-g}L_{tot})}{\partial g_{lm,i}}), 
\end{equation}
which we use in the third term. The result is 
\begin{equation}\label{22}
0 = \partial_k\left[ \frac{(\sqrt{-g}L_{tot})}{\partial q_{,k}} q_{,i} 
 + 
\frac{(\sqrt{-g}L_{tot})}{\partial g_{lm,k}} g_{lm,i} - \delta^k_i 
\sqrt{-g} L_{tot}\right]. 
\end{equation} 
If we consider again the case where the matter Lagrangian does not depend on 
the metric derivatives, we can split the conservation law in the following 
way 
\begin{equation}\label{23}
0 = \partial_k\left[ \sqrt{-g} (\tau^k_{\ i} + t^k_{\ i})\right], 
\end{equation} 
with the canonical matter stress-energy tensor (\ref{6}) and with  
\begin{equation} \label{24}
t^k_{\ i} = 
\frac{\partial L_{grav}}{\partial g_{lm,k}}\ g_{lm,i} - \delta^k_i L_{grav}. 
\end{equation}
Consequently,  we find that the total momentum
\begin{equation} \label{25}
\Pi_i = \pi_i + \int \sqrt{-g}\  t^0_{\ i} \de^3 x 
\end{equation}
is conserved. 
 
We should also mention the fact that we have slightly simplified 
the above considerations by considering the $L_{grav}$ not to depend 
on the second derivatives of the metric. Thus, in all our expressions, 
it is supposed that we omit, in $\sqrt{-g}L_{grav}$ the divergence 
term containing the second derivatives. If this is not done, i.e., 
if one wishes to work with the full Lagrangian $\sim \hat R$, (or, if 
one considers higher order gravity with terms like $\hat R^2 $ etc.) 
then 
one has to take into account additional terms in the field equations 
(\ref{21}), 
as well as in the definition of the tensor (\ref{24}). 
The conservation law (\ref{23}) 
will still be valid and the resulting momentum is of course the same, 
since the surface term is physically irrelevant. The expressions for 
the Euler-Lagrange equations and the canonical stress-energy tensor 
for Lagrangians containing second derivatives of the field variables 
can be found, e.g., in Ref. \cite{jackiw}.

As we can see, there is really nothing obscure in the stress-energy of 
the gravitational 
field. The quantities  $t^{ik}$ are certainly dependent on the coordinate 
system and the conservation law (\ref{23}) 
is not in an explicitely covariant form, but this is a general 
feature of canonical stress-energy tensors, gravitational or not. 
The consequences of this and the 
interpretation in curved spacetime can be found in textbooks on general
relativity. 

We should however point out an important difference. It is not hard to 
argue that $\Pi_i$ in (\ref{25}) 
is actually not only constant, but it is zero.
This is again a result of the fact that (\ref{23}) 
has to be an identity, 
since all the equations of motion (including for the gravitational field) 
 have already been used, and no further constraint can arise on the fields. 
Thus, the integrand of (\ref{23}) is again a relocalization term, and 
the integral over whole space vanishes. In other words, the 
total stress-energy $\tau^i_{\ k} + t^i_{\ k}$ can only be used to define 
the  energy locally, for some finite region of space. The situation, 
under this aspect, is no different from the conventional Landau-Lifshitz 
approach. On the other hand, the total momentum, 
integrated over whole space and 
containing all the fields, is not measurable anyway, because there is nothing 
left with respect to which an eventual variation  could be measured. 
Therefore, the value zero is as good as any other, and certainly better then 
the infinite values obtained, e.g., for the energy,  
in other theories, when gravity is not taken into account.

The tensor (\ref{24}) and its relation to the Landau-Lifshitz tensor 
has also been discussed in Refs. \cite{6} and \cite{7}. The Belinfante 
symmetrization of the tensor (\ref{24}) has been carried out in 
Ref. \cite{jackiw}. 
A  different   
approach to the stress-energy of gravity can be found in Ref. \cite{8}. 
We would also like to point out an interesting paper by Padmanabhan\cite{9}, 
dealing with the question whether general relativity can be 
obtained from a consistent coupling of the spin 2 field to the total 
stress-energy tensor, including its own contributions. Many features 
of the stress-energy concept for gravity and matter are also discussed  
in the textbook Ref. \cite{9a} as well as in Ref. \cite{mont2}.

The reason for including this section is the rather 
 strange fact that, on one hand, in texts on quantum field 
theory, 
field Hamiltonians that correspond to the time component of the canonical 
stress-energy tensor (which is neither a covariant object, nor gauge 
invariant) are widely used, without even mentioning the relevant problems, 
while in 
general relativity, the concept of energy is often considered, for just those 
 reasons,  to be 
badly defined, and in some texts, the discussion is avoided completely.  
Hopefully, it became clear that, concerning momentum conservation,  
their are actually more similarities 
than differences between gravitational and non-gravitational fields.

\section{Poincar\'e gauge theory } 

It is well known that there are no finite dimensional spinor representations 
of the general linear group, and therefore, the introduction of spinor fields 
into the framework of gravitational theory is necessarily accompanied by the 
introduction of a flat tangent space endowed with the Lorentz metric 
$\eta_{ab}$. The relation between physical spacetime and tangent space 
is assured by the existence of a tetrad field $e^a_m$, which allows 
for the introduction of the \textit{curved} Dirac matrices $\gamma_i 
= e^a_i \gamma_a $, where $\gamma_a $ are the usual, constant Dirac matrices. 
The spinors are then considered to be invariant under spacetime
transformations. Local Lorentz invariance is achieved by introducing a 
Lorentz connection. As in general relativity, one has the choice of 
considering the connection either as fundamental field variable, independent 
of the tetrad field, or as a function of the tetrad and its derivatives. 
The first way, which seems to us more satisfying and which we treat in 
this section, is the conception of 
Poincar\'e gauge theory, which is essentially a description of gravity 
in terms of Riemann-Cartan geometry, while the second way is merely  a 
reformulation of general relativity, replacing $g_{ik}$ by $e^a_i$ in 
order to allow for spinor fields, with the disadvantage of having    
derivatives of the gravitational field $e^a_i$ coupling directly to the 
matter fields. This approach will be considered in the next section.

Let us first give a short  review of 
the basic concepts of Riemann-Cartan geometry and
fix our notations and conventions.  
 For a complete introduction 
into the subject, the reader may consult Refs. 
\cite{10} and \cite{11}. 

Latin letters from 
the beginning of the alphabet ($a,b,c\dots $) run from 0 to 3 and 
are (flat) tangent space indices. Especially, $\eta_{ab}$ is the 
Minkowski metric $diag(1,-1,-1,-1)$ in tangent space. Latin letters 
from the middle of the alphabet ($i,j,k \dots $) are indices in a curved 
spacetime with metric $g_{ik}$ as before. We introduce the Poincar\'e gauge 
fields, the tetrad  $e^a_m $ and the connection 
$\Gamma^{ab}_{\ \ m}$ (antisymmetric in $ab$), as well as 
the corresponding field 
strengths, the curvature and torsion tensors
\begin{eqnarray}\label{26}
R^{ab}_{\ \ lm} &=& \Gamma^{ab}_{\ \ m,l} - \Gamma^{ab}_{\ \ l,m} 
                     + \Gamma^a_{\ cl}\Gamma^{cb}_{\ \ m}   
- \Gamma^a_{\ cm}\Gamma^{cb}_{\ \ l}  \\ \label{27}
T^a_{\ lm} &=& e^a_{m,l} - e^a_{l,m} + e^b_m \Gamma^a_{\ bl}
- e^b_l \Gamma^a_{\ bm}. 
\end{eqnarray}
The spacetime connection $\Gamma^i_{ml} $ and the spacetime metric $g_{ik}$
can now be defined through 
\begin{eqnarray}\label{28}
e^a_{m,l} + \Gamma^a_{\ bl} e^b_m &=& e^a_i \Gamma^i_{ml}\\ \label{29}
e^a_ie^b_k \eta_{ab} &=& g_{ik}.  
\end{eqnarray}
It is understood that there exists an inverse to the tetrad, such that 
$e^a_i e_b^i = \delta^a_b $. It can easily be shown that the connection 
splits into two parts, 
\begin{equation}\label{30}
\Gamma^{ab}_{\ \ m} = \hat \Gamma^{ab}_{\ \ m} +
K^{ab}_{\ \ m},
\end{equation} 
such that $\hat \Gamma^{ab}_{\ \ m}$ is torsion-free 
and is essentially a function of $e^a_m$. $K^{ab}_{\ \ m}$ is 
 the contortion tensor (see below). Especially, 
the spacetime connection $\hat \Gamma^i_{ml}$ constructed from 
\begin{equation}\label{31}
e^a_{m,l} + \hat \Gamma^a_{\ bl} e^b_m = e^a_i \hat \Gamma^i_{ml}
\end{equation}
is just the (symmetric) 
Christoffel connection of general relativity, a function of 
the metric only. 

The gauge fields $e^a_m $ and $\Gamma^{ab}_{\ \ m} $ are one-forms, i.e., 
covectors  with respect to the spacetime index $m$. Under a local 
Lorentz transformation (more precisely, the Lorentz part of a 
Poincar\'e transformation, see Refs. \cite{12} and \cite{13}) 
in tangent space, $\Lambda^a_{\ b}(x^m)$, 
they transform as 
\begin{equation}\label{32}
e^a_m \rightarrow \Lambda^a_{\ b}e^b_m, \ \ \ \Gamma^a_{\ bm} \rightarrow 
\Lambda^a_{\ c}\Lambda^{\ d}_{b} \Gamma^c_{\ dm} - 
\Lambda^a_{\ c,m} \Lambda^{\ c}_{b}. 
\end{equation}
The torsion and curvature are Lorentz tensors with respect to their 
tangent space indices as is easily shown. The 
contortion $K^{ab}_{\ \ m}$ is also a Lorentz tensor and is related to 
the torsion through $K^i_{\ lm}= \frac{1}{2}(T_{l\ m}^{\ i}+ T_{m\ l}^{\ i}
- T^i_{\ \ lm})$, 
with $K^i_{\ lm} = e^i_a e^{}_{lb} K^{ab}_{\ \ m}$ and analogously 
for $T^i_{ \ lm}$. The inverse relation is $T^i_{\ lm} = 
-2 K^i_{\ [lm]}$. 

All quantities constructed from the torsion-free connection 
$\hat \Gamma^{ab}_{\ \  m}$ or $\hat \Gamma^i_{lm}$ will be denoted with 
a hat, for instance $\hat R^{il}_{\ \ km}
= e_a^i e_b^l \hat R^{ab}_{\ \ km}$ is the usual Riemann curvature tensor. 
This is consistent with the notation of the previous  sections. 

The gravitational Lagrangian is now constructed using terms at most 
quadratic in curvature and torsion, containing thus no second derivatives 
of the gravitational fields. The most simple candidate is the 
Einstein-Cartan Lagrangian $L_{grav} = - (1/2) R$, with $R 
= e^i_a e^k_b R^{ab}_{\ \ ik}$, which leads essentially back to 
general relativity, with an additional spin-self interaction for spinor 
fields due to a non-dynamical torsion field\cite{10}. 

First, we define the gravitational stress-energy tensor 
as well as the  
spin density tensor in the following way 
\begin{equation}\label{33}
T^i_{\ a} = - \frac{1}{e} \frac{\delta (e  L)}{\delta e^a_i}\ \ \
\text{and}\ \ \ \sigma_{ab}^{\ \ m} = \frac{1}{e} \frac{\delta (e L)}
{\delta  \Gamma^{ab}_{\ \ m}},   
\end{equation}
where $L$ is again the matter Lagrangian, with $S = \int e L \de^4 x$, 
where $e = \det (e^a_i) $. 

We now derive the conservation law that results from the local Lorentz 
invariance. Consider an infinitesimal transformation 
(\ref{32}), with 
 $\Lambda^a_{\ b} = \delta^a_b + \epsilon^a_{\ b}$ 
($\epsilon^{ab}=-\epsilon^{ba}$).  
The  fields $e_m^a$ and $\Gamma^{ab}_{\ \ m}$ undergo 
the following change:
\begin{equation}\label{34}
\delta \Gamma^{ab}_{\ \ m} = - \epsilon^{ab}_{\ \ ,m}- \Gamma^a_{\ cm}
\epsilon^{cb} - \Gamma^b_{\ cm} \epsilon^{ac}\ \ \ \text{and}\ \ \  
\delta e^a_m = \epsilon^a_{\ c} e^c_m. 
\end{equation}
The first equation can be written in the short form $\delta \Gamma^{ab}_{\ \
  m}  = - \De_m \epsilon^{ab}$. 
The change in the matter Lagrangian therefore reads
\begin{equation}\label{35}
\delta (e L) = \frac{\delta (e L)}
{\delta e^a_m} \delta e^a_m 
+ \frac{\delta (e L)}{\delta \Gamma^{ab}_{\ \ m}}
\delta \Gamma^{ab}_{\ \ m}= (T^{[ac]}+ \De_m \sigma^{acm})\epsilon_{ac},  
 \end{equation}
where we have omitted a total divergence. The covariant 
derivative operator $\De_m$ is defined to act with $\Gamma^{ab}_{\ \ m}$ 
on tangent space indices and with $\hat \Gamma^l_{ki}$ (torsion-free) 
on spacetime indices. The requirement of Lorentz gauge invariance 
therefore leads to 
\begin{equation} \label{36}
T^{[ac]}+ \De_m \sigma^{acm}= 0. 
\end{equation}
Slightly more complicated is the case of the coordinate invariance. 
Under an infinitesimal coordinate transformation, 
\begin{equation}\label{37}
\tilde x^i = x^i + \xi^i, 
\end{equation}
the fields $\Gamma^{ab}_{\ \ m}$ and $e^a_m$ transform as spacetime covectors
(or one-forms), i.e., 
\begin{equation}\label{38}
\tilde e^a_m(\tilde x) = e^a_m(x) - \xi^k_{\ ,m} e^a_k 
\ \ \ \text{and}\ \ \ \tilde \Gamma^{ab}_{\ \ m}(\tilde x) = 
\Gamma^{ab}_{\ \  m}(x) - \xi^k_{\ ,m}\Gamma^{ab}_{\ \ k}. 
\end{equation}
Since we are interested in the change of the Lagrangian under an 
active transformation, we have to  evaluate the change of the fields  
at the same point $x$, and thus  have to express the transformed fields 
in the old coordinates, i.e., 
\begin{equation}\label{39}
\tilde e^a_m (x) = \tilde e^a_m(\tilde x) - \tilde e^a_{m,k}(\tilde x) 
\xi^k,\ \ 
\tilde \Gamma^{ab}_{\ \ m}(x) = \tilde \Gamma^{ab}_{\ \ m} (\tilde x)-
\xi^k \tilde \Gamma^{ab}_{\ \ m,k}(\tilde x). 
\end{equation}
In the $\xi$-terms, we can replace $\tilde e^m_a(\tilde x)$ by $e^m_a(x)$ and  
$\tilde\Gamma^{ab}_{\ \ m}(\tilde x)$ by $\Gamma^{ab}_{\ \ m}(x)$, since the 
difference will be of order $\xi^2$. 
Finally, we find
\begin{eqnarray}\label{40}
\delta e^a_m &=&
 \tilde e^a_m(x) - e^m_a(x) = - \xi^k_{\ ,m}e^a_k- \xi^k e^a_{m,k}, 
\\\label{41}
\delta \Gamma^{ab}_{\ \ m} &=& \tilde \Gamma^{ab}_{\ \ m}(x) - 
\Gamma^{ab}_{\ \ m}(x) = - \xi^k_{\ ,m}\Gamma^{ab}_{\ \ k}- \xi^k
 \Gamma^{ab}_{\ \ m,k}.  
\end{eqnarray}
The change of the Lagrangian reads 
\begin{eqnarray}\label{42}
  \delta (e L) &=& \frac{\delta (e L)}{\delta e^a_m} \delta e^a_m 
+ \frac{\delta (e L)}{\delta \Gamma^{ab}_{\ \ m}} \delta 
\Gamma^{ab}_{\  \ m} \nonumber \\  &=& [  
(eT^m_{\ a}e^a_k)_{,m} - eT^m_{\ a}e^a_{m,k}
- (e\sigma_{ab}^{\ \ m}\Gamma^{ab}_{\ \ k})_{,m}  
+  e\sigma_{ab}^{\ \ m}
\Gamma^{ab}_{\ \ m,k} ]\xi^k. 
\end{eqnarray}
Requiring $\delta (e L) = 0$ for arbitrary $\xi^i$ and regrouping 
carefully the terms, we finally get 
\begin{equation}\label{43}
(\De_m T^m_{\ \ b}) e^b_k + T^m_{\ \ b} T^b_{\ mk} 
- R^{ab}_{\ \ mk} 
\sigma_{ab}^{\ \ m} - \Gamma^{ab}_{\ \ k} (\De_m\sigma_{ab}^{\ \ m}
+  T_{[ab]}) = 0. 
\end{equation}
The last term vanishes in view of  
(\ref{36}), such that we find the covariant 
conservation law 
\begin{equation}\label{44}
(\De_m T^m_{\ \ b}) e^b_k + T^m_{\ \ b} T^b_{\ mk} 
- R^{ab}_{\ \ mk} 
\sigma_{ab}^{\ \ m} = 0. 
\end{equation}
In terms of the usual covariant derivative with the Christoffel 
symbols (denoted, as before, with a semicolon), we can 
alternatively write 
\begin{equation}\label{45}
T^{m}_{\ k;m} - K^{im}_{\ \ k} T_{mi} = R^{ab}_{\ \
  mk}\sigma_{ab}^{\ \ m}. 
\end{equation}
It is not hard to show that for matter fields that do not couple 
to $\Gamma^{ab}_{\ \ i}$, and  couple to $e^a_m$  only 
via the metric $g_{ik}$,  $T^i_{\ a}$ defined by 
(\ref{33}) is 
identical to the metric tensor (\ref{2}) (and thus symmetric), while 
$\sigma_{ab}^{\  \ m}$ 
is zero. Then, 
(\ref{45}) reduces immediately to (\ref{5}) (since $K^{im}_{\ \ k}$ 
is antisymmetric in $im$). (It should be clear that, when we speak 
of the symmetry of a quantity like $T^i_{\ a}$, this refers of course 
to the corresponding tensor $T^{ik} = g^{km} (T^i_{\ a} e^a_m)$.) 

Clearly, (\ref{44}) is not a conservation law in the strict sense and gives 
rise to a conserved momentum only for vanishing curvature and torsion. 
Further, just as was the case 
during the derivation of relation (\ref{5}), in order to 
derive (\ref{36}) and (\ref{44}), the matter field equations have to be  
 fulfilled. If this is not the case, we would find additional 
variations $(\delta (eL) / \delta q) \delta q $ in 
(\ref{35}) and (\ref{42}). 
One should have this in mind when splitting the matter Lagrangian in free 
parts and interaction parts. 

We now turn to the canonical stress-energy tensor. 
Since the gravitational 
stress-energy (\ref{33}) is defined by variation with respect to $e^a_m$, 
one might consider $e^a_m$ as the true gravitational field, while 
$\Gamma^{ab}_{\ \ m}$ is an additional matter field. From this point of 
view, one would then define the canonical stress-energy tensor 
(\ref{6}) with 
$q$ containing, apart from $A_i, \psi, \bar \psi, \phi \dots$, also 
the Lorentz gauge field $\Gamma^{ab}_{\ \ m}$. This interpretation is 
 possible, 
since what we usually understand as gravity 
is in fact completely  determined by the geodesic structure of spacetime, 
and thus by the metric, i.e., by $e^a_i$. In this sense, $\Gamma^{ab}_{\ \ m}$
 is an additional matter (gauge) field, leading to new interactions (spin 
precession etc.). However, it is customary to consider the set $(e^a_m,
\Gamma^{ab}_{\ \ m})$ as gravitational fields for the following reason. The 
Lorentz connection is not really the  gauge field  corresponding to the 
Lorentz group, but rather a part of the gauge fields $(\Gamma^{ab}_{\ \ m}, 
\Gamma^a_{\ m})$ corresponding to the Poincar\'e group. Although the 
translational symmetry has been broken by setting $\Gamma^a_{\ m} = e^a_m$ 
(see Refs. 
\cite{12} and \cite{13}  for details), the original Poincar\'e 
structure is still visible in the fact that with each Lorentz transformation, 
apart from the Lorentz connection, we also have to transform $e^a_m$ (see 
(\ref{32})). This reflects the fact that the Poincar\'e group is not simply 
the direct product of Lorentz and translational group. From this aspect, 
it is preferable to consider both the tetrad and the connection as 
gravitational fields and thus, the canonical stress-energy tensor 
of the matter fields $\tau^i_{\ k}$ is to be defined as before by 
(\ref{6}), 
with $q$ containing everything but those fields. 

Then, we proceed as in (\ref{9}), namely 
\begin{equation}\label{46}
\frac{\partial}{\partial x^i} (e L ) = 
\frac{\partial(e L)}{\partial q} q_{,i} 
+ \frac{\partial(e L)}{\partial q_{,k}} q_{,k,i} 
+ \frac{\partial(e L)}{\partial e^a_{m}} e^a_{m,i}
+ \frac{\partial(e L)}{\partial \Gamma^{ab}_{\ \ m}} \Gamma^{ab}_{\ \ m,i}. 
\end{equation} 
It is needless to say that we suppose that the matter fields do not 
couple to the derivatives of $e^a_m$ or $\Gamma^{ab}_{\ \ m}$. 
After all, this 
is  one of the chief reasons for introducing $\Gamma^{ab}_{\ \ m}$
as independent field! Next, we use the matter field equations 
(\ref{8}) (with 
$\sqrt{-g}$ replaced with $e$) to find 
\begin{equation}\label{47}
0 = \partial_k[ e \tau^k_{\ i}] + e \sigma_{ab}^{\ \ m} \Gamma^{ab}_{\ \ m,i}
- e T^m_{\ a} e^a_{m,i}. 
\end{equation}
This is the conservation law that replaces (\ref{11}). Again, 
in the flat limit 
(vanishing curvature and torsion), we find the usual conservation law for 
the canonical stress-energy tensor in special relativity.  
Just as (\ref{11}), this relation is not 
explicitely covariant (neither is $\tau^k_{\ i}$) 
and moreover, it is not even explicitely Lorentz gauge invariant.
(Nevertheless, the relation is valid in any coordinate system and in 
any gauge.)   
This is of course the 
result of the fact that the free gravitational fields are not contained 
in $\tau^k_{\ i}$. (Similarly, as we will see in the next sections, 
in special relativistic field theory, when we consider the Dirac 
field in an electromagnetic background field and define the canonical 
stress-energy tensor without taking into account the free Maxwell field, 
 the resulting tensor will not be $U(1)$ invariant. We stress this 
analogy, just 
in case someone might again begin to  have  doubts on the compatibility  
of the 
concepts of gravity and field energy.) 

The fact that neither (\ref{44}), nor (\ref{47}) lead to a conserved momentum 
vector, does not bother us here. Clearly, as long as, say,  a particle, 
is subjected to gravitational fields, it will not possess a constant 
momentum vector. However, it should be noted that, in contrast to 
general relativity, we cannot, locally, transform the gravitational field to 
zero. Although it is possible (by a combination of 
both gauge and coordinate transformations) to achieve (at some point) 
$e^a_i = \delta^a_i,\  \Gamma^{ab}_{\ \ i} = 0$ 
(see Ref. \cite{11}), it is not generally 
possible to transform away the connection derivatives. Therefore, we cannot 
get rid of the second term in (\ref{47}), which is the analogue of a tidal 
force term (with the particle's intrinsic spin instead of the 
restframe angular momentum of an extended body).    

What we are interested in is the relation between the 
canonical and the gravitational stress-energy tensor. We rewrite 
(\ref{47}) in 
the form 
\begin{equation}\label{48}
0 = \partial_k\left[ e(\tau^k_{\ i} - T^k_{\ i})\right] 
+ (eT^k_{\ i})_{,k} + e 
\sigma_{ab}^{\ \ m} \Gamma^{ab}_{\ \ m,i} - e T^m_{\ a} e^a_{m,i}. 
\end{equation}
By a convenient reordering of the last three terms (i.e., completing 
$\Gamma^{ab}_{\ \ m,i}$ to $R^{ab}_{\ \ im}$, $e^a_{m,i}$ to $T^a_{\ im}$ 
and $(eT^k_{\ i})_{,k} $ to $ e (\De_k T^k_{\ a}) e^a_i$), and using 
the relations (\ref{36}) and (\ref{44}), we find 
\begin{equation}\label{49}
0 = \partial_k\left[ e(\tau^k_{\ i} - T^k_{\ i} - \sigma_{ab}^{\ \ k} 
\Gamma^{ab}_{\ \ i})\right]. 
\end{equation}
As was the case with (\ref{13}),  this is not a conservation law, but an 
identically satisfied relation between the canonical and the gravitational 
 stress-energy tensors. We see that only for $\Gamma^{ab}_{\ \ i} = 0$, 
the momentum vectors $P_i = \int e T^0_{\ i} \de^3 x $ and 
$\pi_i = \int e \tau^0_{\ i} \de^3 x$ will be the same 
(the integration  constant can be shown to be zero again). 
In general field configurations, this will not be the case, except for 
particles with $\sigma_{ab}^{\ \ i} = 0$, i.e., bosonic matter.

It is interesting to remark that in teleparallel theories (i.e., with 
$\Gamma^{ab}_{\ \ i} = 0$ throughout), both tensors turn out to be
equivalent. In order for $\Gamma^{ab}_{\ \ i}$ to vanish everywhere (at 
least in a certain gauge), Lagrange multipliers have to be used  
  to set $R^{ab}_{\ \ lm} = 0$ (see Ref. \cite{14}). Such theories 
seem to be  consistent (the so called one-parameter teleparallel theory), 
but have some rather unnatural features (see Refs. 
\cite{15} and \cite{16} for 
a detailed discussion). Especially, it does  not look very wise 
first to generalize the framework of general relativity from 
10 fields $g_{ik}$ 
(or 16,  $e^a_m$, if you prefer) to 40 fields $(e^a_m, \Gamma^{ab}_{\ \ m})$ 
and 
then, in a next step, to force the new fields to vanish identically 
(or at least to be of pure gauge form, i.e., the corresponding field 
tensor $R^{ab}_{\ \ lm}$ to vanish).   

We conclude that in the general case, and especially in the case of 
Einstein-Cartan theory, the canonical and the gravitational 
stress-energy tensor are  equivalent only if we neglect the 
gravitational field $\Gamma^{ab}_{\ \ m}$ (flat limit). 

In order to be complete, let us also indicate the canonical stress-energy 
tensor for the gravitational field itself, namely 
\begin{equation}\label{50}
t^k_{\ i} = \frac{\partial L_{grav}}{\partial e^a_{m,k}} \ e^a_{m,i}
+ \frac{\partial L_{grav}}{\partial \Gamma^{ab}_{\ \ m,k}}\  
\Gamma^{ab}_{\ \ m,i} - \delta^k_i L_{grav}, 
\end{equation}
which is a direct generalization of (\ref{24}), and which satisfies the 
conservation law 
\begin{equation} \label{51}
0 = \partial_k \left[e(t^k_{\ i} + \tau^k_{\ i})\right], 
\end{equation}
where $\tau^k_{\ i}$ is the canonical matter stress-energy tensor 
(\ref{6}). 
This relation is straightforwardly derived following the usual pattern. 
Thus, again, the total momentum 
\begin{equation}\label{52}
\Pi_i = \pi_i + \int e\ t^0_{\ i}\ \de^3 x 
\end{equation}
is conserved, where $\pi_i = \int e\ \tau^0_{\ i}\ \de^3 x $. We see that 
the canonical approach has a straightforward generalization to Poincar\'e 
gauge theory, while a corresponding form of the Landau-Lifshitz approach 
(\ref{19}) is not known to us. Especially, the requirement that $t^{ik}$ be 
symmetric now seems unfounded, because $T^{ik}$ (from (\ref{33})) 
is itself asymmetric and moreover, there is no reason, in the presence of 
spinning matter fields,  for the orbital angular momentum to be conserved 
separately. Nevertheless, 
it remains an  interesting  question if a 
relation similar to (\ref{19}), 
i.e., containing $T^{ik}$ instead of $\tau^i_{\ k}$, 
can be found in the framework of Poincar\'e gauge theory. A related
investigation 
will be carried out in the appendix. 

In the special case of Einstein-Cartan theory, where 
$L_{grav}= -\frac{1}{2} R$,  we readily find 
\begin{equation} \label{53}
t^k_{\ i} = - (e^k_a e^m_b) \Gamma^{ab}_{\ \ m,i} - \delta^i_k R, 
\end{equation}
or, after some simple manipulations, 
\begin{equation}\label{54}
t^k_{\ i} = - G^k_{\ i} - (\frac{1}{2} T^k_{\ ab} + \frac{1}{2} 
T^l_{\ [bl}e^k_{a]}) \Gamma^{ab}_{\ \ i} - \frac{1}{e}( e\ e^k_a e^m_b 
\Gamma^{ab}_{\ \ i})_{,m}. 
\end{equation}
We see that, in vacuum, where the torsion vanishes, $t^k_{\ i}$ equals, 
up to a relocalization term, the negative of the Einstein tensor $G^k_{\ i}$, 
which vanishes also. Thus, just as in general relativity, $t^k_{\ i}$ 
contains only surface terms. The same holds for the total stress-energy 
in the presence of matter fields. Especially, the total momentum 
is zero again. 
This is a feature shared with the Landau-Lifshitz approach. 

The momentum conservation equation (\ref{51}) 
leads us to conclude that quite 
generally, the 
canonical stress-energy tensor corresponds to the physical energy,  
stress and momentum densities, while the gravitational tensor plays 
the role as source of gravity. The conclusion is  based on the following. 
First, it is standard procedure to use the  time component of the 
canonical tensor as Hamiltonian in quantum field theory.  
This has not led to any inconsistencies so far. 
Second, the (total) canonical momentum is the most simple conserved 
quantity that can be derived from a general Lagrangian, and it should   
therefore have a fundamental, physical contents. Even for 
exotic theories, such as, e.g., Ref. \cite{19}, based on a  symmetry 
between anti-gravitating and gravitating matter,  
with  anti-gravity 
coupling  with the opposite sign to the Lorentz connection, 
it turns out that, when 
comparing the gravitating with the anti-gravitating matter section, the 
metric stress-energy tensors (i.e., the source of the gravitational field)
are of opposite sign, while this is not the case with the canonical tensor. 
As a result, the energy density and the Hamiltonian as defined from 
the canonical tensor, are still positive, as they should 
(for vacuum stability). 

More generally, energy (momentum) is, by its very definition, 
 canonically  conjugate to  time  (space), i.e., in quantum field 
theory, energy (momentum) corresponds to the generator of 
time (space) translations. This is directly incorporated in the canonical 
definition of the stress-energy tensor, while the variationally 
defined  gravitational tensor is not, a priori, related to such 
concepts. We should, however mention that, apart from their role 
as source of gravity, the  gravitationally defined tensors 
(\ref{33})  play an important role on the kinematical level. 
In particular, the particle momentum (as opposed to the field momentum) 
in a semiclassical picture is directly related to the integrated 
gravitational stress-energy tensor. We refer to Ref. \cite{17} for 
details on particle motion in Poincar\'e gauge theory.

There are of course some well known problems. First, $\tau^i_{\ k}$ is not 
a true tensor. That seems physically acceptable. Energy or momentum densities 
are not observer independent quantities. The introduction of a Hamiltonian 
is necessarily preceeded by a $(3+1)$-split of spacetime, and the
consideration of the other components will break the remaining  
symmetry. Second, the tensor is not Lorentz gauge invariant. Moreover, 
the momentum $\pi_i$ of the matter fields is not gauge invariant. 
Thus, in order to determine the four-momentum of a 
certain matter distribution (in a gravitational background), 
we have to fix the Lorentz gauge. Especially, 
the matter Hamiltonian will depend on the gauge choice. This seems rather 
disturbing, it is, however, quite usual. Exactly the same thing happens 
in the case of a charged field in an electromagnetic background. (The problem  
arises already at the level of quantum mechanics\cite{18}.) Only the 
complete, Maxwell + Dirac Hamiltonian is gauge invariant, whereas the 
Dirac Hamiltonian on the Maxwell background  will badly depend on the gauge 
choice. Thus, once again, nothing special with gravity!

We close our discussion of the stress-energy of 
gravitational fields at this point. 
Before we turn to specific examples (point charge, Dirac-Maxwell system), 
we briefly 
include a section on tetrad gravity, i.e., general relativity expressed 
in terms of $e^a_i$ and coupled to spinor fields.

\section{Tetrad gravity} 

With the background of the last section in mind, it now straightforward 
to include spinor fields into general relativity, without the use of 
an additional, independent Lorentz connection. Just as before, 
we consider a flat tangent space with metric $\eta_{ab}$ and 
the basic gravitational fields are now $e^a_i$, from which we 
define the metric $g_{ik} = e_i^a e_k^b \eta_{ab}$. With the metric, 
we can form the Christoffel symbols $\hat \Gamma^i_{lm}$ and then 
\textit{define} the connection $\hat \Gamma^{ab}_{\ \ m}$ by means of 
(\ref{31}). The connection is automatically  torsion free, i.e., $e^a_{i,k} + 
\hat \Gamma^a_{\ bk}e^b_i - e^a_{k,i} - \hat \Gamma^a_{\ bi}e^b_k 
= e^a_l( \hat \Gamma^l_{ik} - \hat \Gamma^l_{ki}) = 0$ in view of 
the symmetry of the Christoffel connection.

We still require local Lorentz invariance  in the form 
$e^a_i \rightarrow \Lambda^a_{\ b} e^b_i$. Then, $\hat \Gamma^{ab}_{\ \ i}$ 
transforms automatically as Lorentz connection, i.e., in the same way as
$\Gamma^{ab}_{\ \ i} $ in (\ref{32}). 
Local Lorentz gauge invariance of the matter 
field Lagrangian is assured through the minimal coupling prescription 
$\partial_k \rightarrow \partial_k - \frac{i}{4} 
\hat \Gamma^{ab}_{\ \ k} \sigma_{ab}$ when acting on spinors ($\sigma_{ab}$
are the Lorentz generators), i.e., in 
the same way as in Poincar\'e gauge theory, but with  
$\hat \Gamma^{ab}_{\ \  i}$ replacing  $ \Gamma^{ab}_{\ \ i}$.  

Thus, we have necessarily derivatives of $e^a_i$ (contained in $\hat 
\Gamma^{ab}_{\ \  i}$) coupling to the spinor fields, which makes, 
in our view, this approach less attractive then the Poincar\'e gauge 
approach. 

We define the gravitational stress-energy tensor  through 
\begin{equation}\label{55}
T^i_{\ a} = - \frac{1}{e} \frac{\delta (e  L)}{\delta e^a_i}, 
\end{equation}
which is formally the same as (\ref{33}), with the difference that here, 
the variation includes in addition the tetrad field hidden in the 
connection. (Clearly, for bosonic matter, which couples to $e^a_i$ only 
via $g_{ik}$, $T^i_{\ a}$ is identical to the metric tensor 
(\ref{2}), after 
converting the tangent space index into a spacetime index.) 

An interesting feature of (\ref{55}) is its symmetry, even in presence of 
spinor fields. This is the result of the invariance under $\delta e^a_i 
= \epsilon^a_{\ b} e^b_i$ (the infinitesimal form of the Lorentz 
transformation, with antisymmetric $\epsilon^{ab}$). Indeed, under 
such a transformation, assuming again that the matter field equations 
$\delta (eL)/ \delta q = 0 $ are satisfied, we find 
\begin{equation}\label{56}
\delta (e L) = \frac{\delta (eL)}{\delta e^a_i} \delta e^a_i  
= - e T^i_{\ a} e^c_i \epsilon^a_{\ c} = -e T^{[ca]} \epsilon_{ac}, 
\end{equation}
and therefore $T^{[ac]} = 0$. This is the relation corresponding to 
equation (\ref{36}) of the previous section. 

Note that, despite the apparently simple relations, the actual 
evaluation of $T^i_{\ a}$ can be rather involved, because the expression of 
$\hat \Gamma^{ab}_{\ \ i} $ in terms of $e^a_i$ is not really simple 
(just think of all the metric derivatives contained in the right 
hand side of (\ref{31})). Thus, we get a symmetric, but  complicated 
stress-energy tensor (it is actually the Belinfante-Rosenfeldt tensor, 
see Ref. \cite{10}.) 

The symmetry of $T^{ik}$  is also necessary in view of the field 
equations $\hat G^{ik} = T^{ik}$ (recall that we are dealing with general 
relativity again). From this, it also follows that $T^{ik}_{\ \ ;i} = 0$. 
Indeed, under a spacetime transformation, the tetrad transforms as 
before (Eq. (\ref{40})), and the conservation law 
\begin{equation}\label{57}
(e T^i_{\ a} e^a_k)_{,i} + e T^i_{\ a} e^a_{i,k} = 0 
\end{equation}
is easily derived. It is not hard to show that, for $T^{ik}$ symmetric, 
this is equivalent to $T^{ik}_{\ \ ;i} = 0$, as was to be expected. 

Consider now the the derivative of the Lagrangian 
\begin{equation} \label{58}
\partial_i (e L) = \frac{\partial (eL)}{\partial q} q_{,i} 
+\frac{\partial (eL)}{\partial e^a_k} e^a_{k,i} 
+\frac{\partial (eL)}{\partial e^a_{k,l}} e^a_{k,l,i}, 
\end{equation}
use the matter field equations (\ref{8}), the definitions  (\ref{6}) 
and (\ref{55}), 
as well as the relation (\ref{57}) to find  
\begin{equation}\label{59}
0 = \partial_k\left[e (\tau^k_{\ i}- T^k_{\ i} + \frac{\partial L}
{\partial e^a_{m,k}} e^a_{m,i})\right],   
\end{equation}
which is an identically satisfied relation between the canonical and 
the gravitational tensors, the analogue of 
(\ref{17}). It shows that, for 
matter fields coupling to the tetrad derivatives (i.e., spinor matter), 
both tensors are only equivalent in the flat limit $e^a_{m,i} =0$. 

Let us remind the reader once again that the apparent contradiction to 
Ref. \cite{4}, where a similar analysis has been carried out using a spinor 
formalism, is due to the different definition of the canonical tensor, 
which has been defined with covariant derivatives in Ref. 
\cite{4}. From the 
result of those authors, we can therefore not conclude anything concerning 
the behavior of the tensor (\ref{6}) or its relation to $T^{ik}$.   

Let us also note that the construction of the canonical stress-energy 
tensor for the gravitational field in the way of 
(\ref{24}), with $g_{ik}$ 
replaced by $e^a_i$, is not possible in the case of matter fields 
coupling to the tetrad derivatives. Although one can derive the relation 
corresponding to Eq. (\ref{24}), 
it is obviously not possible, in the second term, 
to replace $L_{tot}$ by $L_{grav}$, and thus to split the total stress-energy 
tensor into a matter and a gravitational part. This is one more argument 
in favor of an independent Lorentz connection and the first order formalism 
of Poincar\'e gauge theory.

This concludes our analysis of the relation between the canonical 
and the gravitational stress-energy tensors. Let us stress once again 
that even in those cases, where we have found \textit{equivalence}, 
we have by no means \textit{equality}. Especially, the canonical 
tensor is not even a covariant quantity (i.e., a tensor) and in 
general lacks gauge invariance (see, e.g.,  the Maxwell case). 
It means simply that they both lead to the same momentum vector. 
(Which does still not mean that this momentum vector is conserved!) 

In the next sections, we will consider concrete examples, focusing 
mainly on the time component of the momentum vector (the field Hamiltonian) 
and on the flat limit.

\section{Point charge in general relativity}

As a first, classical example, we consider a charged point mass. Since 
in this non-quantum mechanical approach, the particle is directly described 
in terms of its position and not in terms of field variables, the canonical 
approach is not really accessible, and we confine ourselves to the discussion 
of some features of the metrical stress-energy tensor. 

Consider the point mass action 
\begin{equation}\label{60}
S_m = - \int m \ \de \tau,   
\end{equation}
where $\tau $ is the proper time curve parameter. In order to write this 
in the form $S_m = \int L \sqrt{-g}    \de^4 x$, we take the following 
steps: 
\begin{eqnarray}\label{61}
S_m &=& - \int m \frac{\de \tau}{\de t} \de t \\ \label{62}
&=& - \int m \frac{\de \tau}{\de t}\ \delta^{(3)}(x - x_0) \de^4 x.
\end{eqnarray}
where $x_0 $ is the (time dependent) position of the particle. Now, 
use $\de \tau = \sqrt{g_{ik} \de x^i \de x^k} = \sqrt{g_{ik} v^i v^k} \de t$
with the coordinate velocity $v^i = \de x^i /\de t = (1, \vec v)$. The result 
is 
\begin{equation}\label{63}
S_m = - m \int \sqrt{g_{ik} v^i v^k} \ \delta^{(3)}(x- x_0) \de^4 x, 
\end{equation}
which is, by construction, invariant under coordinate transformations. 
The tensor (\ref{2}) therefore has the form 
\begin{equation}\label{64}
T^{ik}_m = \frac{1}{\sqrt{-g}} \frac{m}{\sqrt{g_{lm}v^l v^m}} 
\ \delta^{(3)}(x-x_0) v^i v^k. 
\end{equation}
With the help of the proper time velocity $u^i = \de x^i / \de \tau$ and 
the parameter normalization $u_i u^i = 1$, we get the alternative form  
\begin{equation}\label{65}
T^{ik}_m = \frac{m}{\sqrt{-g}}\ \delta^{(3)}(x-x_0) u^i v^k. 
\end{equation}
Note that for $\vec v = 0$ (point particle at rest), we get 
\begin{equation}\label{66}
T^{00}_m = \frac{1}{\sqrt{g_{00}}}\frac{m}{\sqrt{-g}}
\ \delta^{(3)}(x-x_0), 
\end{equation}
and $T^{ik}_m = 0 $ for the other components. 
This should thus be the tensor that appears as source term 
of the Einstein equations in order to derive the Schwarzschild solution. 

We wish here to point out several problems with this approach. First 
of all, the expression (\ref{66}) is rather formal, because in practice, 
we know that the metric diverges at the point $x = 0$ (we suppose 
that the particle is located at $x_0 = 0$), which is just 
the point where the stress-energy tensor is supposed to be 
of importance. This problem is, however, 
not hard to cure. One can simply replace the delta function by 
a more general density $\rho(x)$, still normalized by $\int \rho\ \de^3 x =1$. 

More fundamental is the horizon problem. If the particle is still a black 
hole (i.e., if $\rho $ is still quite concentrated around the origin), 
then, we know from the Schwarzschild solution, that at the horizon, 
$g_{00}$ changes sign and becomes negative. This, however, is incompatible 
with the expression (\ref{66}). Indeed, already from (\ref{64}), we see that 
$g_{ik} v^i v^k$ should be positive, and thus, since inside the 
horizon, where the mass distribution is located, $g_{00} $ is negative, 
we cannot have $\vec v = 0$, i.e., the particle cannot be at rest. 
This might make sense for a test particle in the Schwarzschild field, 
but as far as the source itself is concerned, it is rather a contradiction. 

Where is the origin of this problem? Well, it is not hard to see that 
the problem first occurs when we go from 
(\ref{61}) to (\ref{62}). One can then also 
guess the profound reason: At the horizon, $g_{00}$ and $g_{rr}$ change 
signs, which means essentially that $t$ becomes a spacelike and $r$ a 
timelike coordinate. This, however, does not mean that time becomes 
spacelike, but  rather that $r$ takes over the role of the time 
coordinate and $t$ that of  a 
space coordinate. Therefore, when we suppose the particle to be 
described by a density in 3d space, i.e., $\int \delta^{(3)}(x-x_0)  \de^3 x
= 1$, or $\int \rho(x) \de^3 x = 1$, 
where $\de^3 x = \de x^1 \de x^2 \de x^3$, we actually do not integrate 
over space, but over two space and one time dimension. Which means, 
in the case of the  delta function, that the particle exists not at one 
point, but at only one instant! (Or for a  short while, 
in the case of  a more general density.) This is certainly not what we 
intended. 
The argument makes clear that the problem is not caused by the point-nature 
of the particle, but is generally present in expressions like 
(\ref{62}), which are 
not written  in an explicitely  covariant form. Unfortunately, although  
the form (\ref{65}) can be found in many textbooks, see e.g., 
 Ref. \cite{5}, \S 106, the above  problems are usually not mentioned.

However, since this problem is specifically related to gravity, and
we are interested mainly in the flat limit, we will ignore those 
difficulties  and continue with (\ref{64}). 

In the flat limit, $g_{ik} = \eta_{ik}$, the tensor reduces to   
\begin{equation}\label{67}
T^{ik}_m = \frac{m}{\sqrt{1- \vec v^2}} \ \delta^{(3)}(x-x_0) v^i v^k,   
\end{equation}
and especially,
\begin{equation}\label{68}
T^{00}_m = \frac{m}{\sqrt{1- \vec v^2}}\ \delta^{(3)}(x- x_0), 
\end{equation}
which is recognizable as the (special) 
relativistic kinetic energy density of the particle.

Next, consider the action of the free electromagnetic (EM)  field 
\begin{equation}\label{69}
S_{EM} = - \frac{1}{4} \int F^{ik} F_{ik} \sqrt{-g}\ \de^4 x. 
\end{equation}
The stress-energy tensor is found to be  
\begin{equation}\label{70}
T^{ik}_{EM} = - F^{il} F^k_{\ l} + \frac{1}{4} g^{ik} F^{lm}F_{lm}, 
\end{equation}
whose time component, in the flat limit, reduces to the energy density 
of the EM field, 
\begin{equation}\label{71}
T^{00}_{EM} = \frac{1}{2} (\vec E^2 + \vec B^2). 
\end{equation}
Finally, consider the so-called interaction part of the action, 
\begin{equation}\label{72}
S_{int} = - \int e A_i \de x^i.
\end{equation}
As before, this can be transformed 
\begin{equation}\label{73}
S_{int} 
= - \int e A_i v^i \de t = - \int e A_i v^i \ \delta^{(3)} (x - x_0) \de^4 x.
\end{equation}
First, we derive the current density 
\begin{eqnarray}\label{74}
j^i &=& -\frac{1}{\sqrt{-g}} \frac{\delta (L \sqrt{-g})}{\delta A_i}  
\nonumber \\
&=& \frac{1}{\sqrt{-g}}\  e \ v^i \ \delta^{(3)}(x-x_0). 
\end{eqnarray}
We see then, that (\ref{73}) can be written in the form 
\begin{equation}\label{75}
S_{int} = - \int A_i j^i \sqrt{-g} \ \de^4 x,  
\end{equation}
which is the form  usually found in textbooks. 
(Note that the expression (\ref{74}) is also given explicitely in Ref. 
\cite{5}, \S 90, 
  although 
it has similar  problems as (\ref{64}).) 
We should caution that  the form (\ref{75}) is actually  
quite unsuitable for the derivation of the stress-energy tensor. Indeed, 
apparently, the integrand is metric dependent via the 
factor $\sqrt{-g}$. Moreover, someone might come up with the idea that 
one should write $j^i A_i $ as $g^{ik} j_i A_k $  or as $g_{ik} j^kA^i$, 
each of which leads, with (\ref{2}), to a different stress-energy tensor. 

On the other hand, from the explicit form (\ref{73}), 
it is immediately clear that 
the integrand $L_{int} \sqrt{-g}$ is completely independent of the metric 
tensor, and thus 
\begin{equation}\label{76}
T^{ik}_{int} = 0. 
\end{equation} 
Therefore, from the complete action for the point charge in an electromagnetic 
field, 
\begin{eqnarray}\label{77}
S &=& S_m + S_{int} + S_{EM} \nonumber \\
&=&- \int m \ \de \tau   
- \int e A_i \de x^i
- \frac{1}{4} \int F^{ik} F_{ik} \sqrt{-g}\ \de^4 x, 
\end{eqnarray}
we derive the stress-energy tensor 
\begin{equation} \label{78}
T^{ik} = \frac{m}{\sqrt{-g}}\ \delta^{(3)}(x-x_0) u^i v^k 
- F^i_{\ l} F^k_{\ l} + \frac{1}{4} g^{ik} F^{lm}F_{lm},  
\end{equation}
and especially, for the time component in the flat limit 
\begin{equation}\label{79}
T^{00} = 
 \frac{m}{\sqrt{1- \vec v^2}}\ \delta^{(3)}(x- x_0) + 
\frac{1}{2} (\vec E^2 + \vec B^2). 
\end{equation}
Therefore, we find for the conserved energy the well known expression  
\begin{equation}\label{80}
\mathcal E =  \frac{m}{\sqrt{1- \vec v^2}} + 
\int \frac{1}{2} (\vec E^2 + \vec B^2) \de^3 x. 
\end{equation}
Apparently, the energy is  composed only of the kinetic energy 
of the particle and of the photons, without any \textit{interaction} or 
\textit{potential} energy  (like the potential energy of the charge 
in an exterior field, or the self-interaction energy with its own field).   
This, however, is an illusion, since 
 those contributions are very well contained in the second term 
(only the transverse part of the electromagnetic field corresponds to 
the kinetic photon energy). We will derive explicit expressions in the next 
section.

\section{The Dirac particle}

In this section, we are interested in the stress-energy tensors of 
a Dirac particle in an electromagnetic field. Conveniently, we 
work in  the framework of Poincar\'e gauge theory. 

The Dirac Lagrangian  reads\cite{10} 
\begin{equation}\label{81}
L_D = \frac{i}{2} \left[\bar \psi \gamma^m (D_m - ie A_m ) \psi - \bar \psi 
(\overleftarrow D_m + i e A_m) \psi \right] - m \bar \psi \psi, 
\end{equation}
with $\gamma^m = e^m_a \gamma^a $ and 
$D_m = \partial_m - \frac{i}{4} \Gamma^{ab}_{\ \ m} \sigma_{ab}$, where  
$\sigma_{ab} = (i/2) [\gamma^a, \gamma^b]$ are the Lorentz generators. 
This Lagrangian is invariant under a Lorentz gauge transformation 
\begin{equation}\label{82}
\psi \rightarrow e^{-\frac{i}{4} \epsilon^{ab} \sigma_{ab}} \psi, 
\end{equation} 
while $e^a_m $ and $\Gamma^{ab}_{\ \ m}$ undergo the transformation 
(\ref{32})
with $\Lambda^a_{\ b} = \delta^a_b + \epsilon^a_{\ b}$ (infinitesimally). 

We now derive the Dirac equation and take the flat limit ($e^a_m = \delta^a_m, 
\Gamma^{ab}_{\ \ m} = 0)$. The result is 
\begin{equation}\label{83}
i \gamma^m( \partial_m - ie A_m) \psi = m \psi, 
\end{equation}
or, in Schroedinger form, 
\begin{equation}\label{84}
H \psi = (\beta m + 
\vec \alpha \cdot (\vec p + e \vec A) - e  A_0) \psi = i \partial_0 \psi, 
\end{equation}
with $\vec p = - i \vec \nabla,\  \alpha^{\mu} = \beta \gamma^{\mu} 
\ (\mu = 1,2,3)$, and $\beta = \gamma^0$. 
From  (\ref{33}), 
we find for the gravitational stress-energy tensor (taking 
again the flat limit)
\begin{equation}\label{85}
T^i_{D k} = \frac{1}{2}\left[\bar \psi \gamma^i(i \partial_k + e A_k)\psi 
- \bar \psi( i \overleftarrow \partial_k - e A_k) \gamma^i \psi\right]. 
\end{equation}
Note that this tensor is, by itself, $U(1)$  gauge invariant. 
(Actually, if one does not take the 
flat limit, the resulting tensor is also Lorentz gauge invariant.) 
 
If we had started within the framework of general relativity, in the  
 tetrad formalism of section 5, we would have found a tensor 
 containing  additional terms, 
rendering the total 
tensor symmetric\cite{10}. However, as far as the flat limit is 
concerned, both tensors are equivalent (i.e., lead to identical momentum 
vectors) and the differences between both  approaches are irrelevant for the 
argumentation of this section, which focuses on other points. We therefore 
choose to work with Poincar\'e gauge theory, where the irrelevant 
relocalization terms are absent right from the start.

As before, we are interested in the time components. Integrating $T_D^{00} = 
T^0_{D 0}$, we find 
\begin{eqnarray}\label{86}
\int T_D^{00}\de^3 x 
&=& \frac{1}{2} \int \left[\bar \psi \gamma^0 (i \partial_0 + e A_0)
\psi - \bar \psi (\overleftarrow \partial_0 - e A_0) \gamma^0 \psi
\right] \de^3 x
\nonumber \\
&=& 
\frac{1}{2} \int\left[ \psi^{\dagger} 
(i \partial_0 + e A_0)\psi - \psi^{\dagger} (\overleftarrow 
\partial_0 - e A_0) \psi\ \right ]\de^3 x  \nonumber \\
&=& \int \psi^{\dagger} 
(i \partial_0 + e A_0) \psi\ \de^3 x- \frac{\de }{\de t} \int \psi^{\dagger} 
\psi \ \de^3 x.  
\end{eqnarray}
The last term vanishes in view of the conservation law 
$(\bar \psi \gamma^m \psi)_{,m} = 0$, which follows from the Dirac equation  
(related directly to charge conservation). Therefore, using 
(\ref{84}), we 
can write alternatively 
\begin{equation}\label{87}
\int T_D^{00} \de^3 x = \int \psi^{\dagger} (H + e A_0) \psi\ \de^3 x.  
\end{equation}
Thus, apparently, the energy is related to the operator 
$H + e A_0 = \beta m + \vec \alpha \cdot (\vec p+ e \vec A) $.  
Recall that the operator $\vec p + e \vec A $ (and not $\vec p$) 
is the kinetic momentum that reduces, in the classical limit, to 
$m \vec v$. This is clear anyway if one considers the static case, where 
$\vec A $ is related to the magnetic field, which, as is well known, does 
not change the particle's energy on a classical level since the force induced 
by it is perpendicular to the velocity. Thus, apart from 
purely quantum mechanical contributions (like the $\vec \sigma \cdot \vec B$ 
term contained implicitly in (\ref{84})), 
the major, classical, interaction energy, namely the 
potential energy $- e A_0$ of the electron in the electric field, 
is not contained in the expression (\ref{87}). 
This is in complete analogy with 
the classical case, namely the result (\ref{68}) 
(together with 
the fact that there was no contribution from the interaction part, see 
(\ref{76})). 

Adding the stress energy-tensor (\ref{70}) of the EM field (it is not 
hard to show that the same tensor emerges from a variation with 
respect to $e^a_i$ instead of $g_{ik}$), we find 
for the total energy 
\begin{eqnarray}\label{88}
\mathcal  E &=& \int (T_D^{00} + T_{EM}^{00}) \de^3 x 
\nonumber \\ &=& \int \psi^{\dagger} 
(H + e A_0) \psi\ \de^3 x + \int \frac{1}{2}
(\vec B^2 + \vec E^2 ) \ \de^3 x  \nonumber \\
&=&
\int \psi^{\dagger} (\beta m + \vec \alpha \cdot (\vec p + e \vec A))\psi 
+ 
\int \frac{1}{2}
(\vec B^2 + \vec E^2 ) \ \de^3 x,  
\end{eqnarray}
 and for the energy-density  
\begin{equation}\label{89}
 E  = \psi^{\dagger} (H + e A_0)\psi  + \frac{1}{2} (\vec B^2 + \vec E^2)  
= \psi^{\dagger}(\beta m + \vec \alpha \cdot (\vec p + e \vec A))\psi 
+ \frac{1}{2} (\vec B^2 + \vec E^2). 
\end{equation}
These expressions are to be compared with those in 
(\ref{79}) and (\ref{80}). 
If the Lagrangian is supposed to be complete (i.e., if all the 
matter field equations are satisfied), then this expression 
for the energy is conserved, in view of 
(\ref{44}), which reduces to 
$T^i_{\ k,i} = 0$ in the flat limit. However, if  there is, say,  
an additional (exterior) electromagnetic field (considered not 
to be influenced by the electron), then we cannot guarantee for 
anything.  

The canonical approach, as far as the Dirac particle is concerned, 
is probably more familiar to most physicists. The corresponding 
tensor  has the form 
\begin{eqnarray}\label{90}
\tau^i_{D k} &=& \frac{\delta  L_D}{\delta \psi_{,i}} \psi_{,k} 
+ \bar \psi_{,k} \frac{\delta  L_D }{\delta \bar \psi_{,i}} - 
\delta^i_k  L_D \nonumber \\
&=& \frac{1}{2} (\bar \psi \gamma^i (i \partial_k)\psi - \bar \psi 
(i \overleftarrow \partial_k) \gamma^i \psi) 
\end{eqnarray}
for the Dirac particle (with $L_D$ from 
(\ref{81}), taking the flat limit), 
and  
\begin{eqnarray}\label{91}
\tau^i_{EM k} &=& \frac{\delta  L_{EM}}{\delta A_{m,i}} A_{m,k} 
- \delta^i_k  L_{EM} \nonumber \\
&=& F^{mi} A_{m,k} + \frac{1}{4} \delta^i_k F^{lm}F_{lm}
\end{eqnarray}
for the Maxwell field. 
Let us begin with expression 
(\ref{90}). From the time component, using (\ref{84}) 
and partially  integrating, we readily find 
\begin{equation}\label{92}
\mathcal E_D = \int \tau_D^{00} \de^3 x = \int \psi^{\dagger} H \psi\ \de^3 x, 
\end{equation}
a relation that can be found in any textbook.

This expression, as opposed to its gravitational counterpart 
(\ref{87}) is not $U(1)$
gauge invariant (and it would also not be Lorentz gauge invariant, 
even if we had not  taken the flat limit). 

As to (\ref{91}), some authors seem to believe that it is equivalent 
(i.e., related by a relocalization) to its symmetric counterpart 
(\ref{70}). 
This, however, is not true in general. Indeed, from (\ref{91}), we find  
\begin{eqnarray}\label{93}
\tau^i_{EM k} &=& -F^{mi}F_{mk} + \frac{1}{4} \delta^i_k F^{lm}F_{lm}  
 + F^{mi}A_{k,m} \nonumber \\
&=& T^i_{EMk}  + (F^{mi}A_k)_{,m} - F^{mi}_{\ \ ,m} A_k 
\end{eqnarray}
The second term is indeed a relocalization term.  The last 
term, however, does not vanish in presence of a source term for the
electromagnetic fields. Indeed, the field equations have the form 
\begin{equation}\label{94}
F^{mi}_{\ \ ,m} = j^i, 
\end{equation}
or, for the explicit case of the Dirac particle, 
\begin{equation}\label{95}
F^{mi}_{\ \ ,m} = - e \bar \psi \gamma^i \psi. 
\end{equation}
Therefore, we find for the time component of (\ref{91})  
\begin{equation}\label{96}
\int \tau_{EM}^{00} \de^3 x 
= \int \left[\frac{1}{2}(\vec B^2 + \vec E^2) + e \psi^{\dagger} 
\psi A_0\right] \de^3 x. 
\end{equation}
Taking the sum of (\ref{92}) and (\ref{96}), we find for the total energy 
\begin{equation}\label{97}
\mathcal E = \int \left[\psi^{\dagger} H \psi + e \psi^{\dagger} A_0 \psi 
+ \frac{1}{2}(\vec B^2 + \vec E^2)\right]\de^3 x, 
\end{equation}
which is exactly the same expression as derived from the gravitational 
tensor (see (\ref{88})). This confirms the result (\ref{49}), 
which states that 
in the flat limit, $T^i_{\ k}$ and $\tau^i_{\ k}$ are equivalent. 

The main point we wish to stress in this section is the fact that, 
if one considers only parts of the Lagrangian, i.e., if one splits into 
free and interaction contributions, then the equivalence does not hold 
anymore. The term $ \int \psi^{\dagger} e A_0 \psi \de^3 x$ in 
(\ref{97}), which is 
clearly an interaction (or self-interaction) term,  was found, in the 
gravitational approach, 
to be contained in the Dirac part of the stress-energy tensor 
(in a certain sense, in the Dirac field Hamiltonian), while in the canonical 
approach, the same term was contained in the electromagnetic field 
contributions (i.e., in the Maxwell Hamiltonian). 

This is an important point, especially if one is to consider a Dirac  
particle in a background field (e.g., to evaluate atomic spectra), or 
alternatively, a photon traveling through a background distribution 
of electrons. In each case, one will omit a part of the Lagrangian, and 
one will have to take care to use the appropriate stress-energy tensor.

Let us now take a closer look at the term $\int e \psi^{\dagger} 
\psi A_0 \de^3 x$. 
If we use the Coulomb gauge $\vec \nabla \cdot \vec A = 0$, we can solve 
(\ref{94}) for $A_0 $ in the form 
\begin{equation}\label{98}
A_0(x) = \frac{-e}{4\pi} \int \frac{\psi^{\dagger} (x') \psi(x')}{|x-x'|}
\ \de^3 x', 
\end{equation}
and we can write (\ref{96}) as 
\begin{equation}\label{99}
\int \tau_{EM}^{00} \de^3 x = \int \frac{1}{2}(\vec B^2 + \vec E^2)\de^3 x - 
\frac{e^2}{4 \pi} 
\int \ \frac{\psi^{\dagger}(x) \psi(x) \psi^{\dagger}(x') 
\psi(x')}{|x-x'|}\ \de^3 x'\ \de^3 x.
\end{equation}
The last term is the well known instantaneous four-fermion Coulomb term. 
If we split the electric field into a longitudinal and a transverse 
component, $\vec E = \vec E_T + \vec E_L$, with $\vec \nabla E_L = \rho = j^0$ 
and $\vec \nabla E_T = 0$, then we recognize in (\ref{99}) the 
well-know Hamiltonian used in QED in the form 
\begin{equation}\label{100}
\int \tau_{EM}^{00} \de^3 x = \int \frac{1}{2}(\vec B^2 + \vec E^2_T) \de^3 x, 
\end{equation}
i.e., the non-propagating, longitudinal components of $\vec E$ 
 are not contained in the electromagnetic part of the  
canonical stress-energy tensor. (The 
mixed terms $\vec E_T \cdot \vec E_L$ contained in $\vec E^2$ lead 
only to a surface term, while the longitudinal terms $\vec E_L^2$ 
just cancel against  the four-fermion term in (\ref{99}).)   
However, the same term  appears again in  the Dirac part of the canonical 
tensor (\ref{92}) (see $H$ in (\ref{84})). 
As a result, the total energy (\ref{97}), which 
is the same in both canonical and gravitational approaches, can be written
as 
\begin{eqnarray}\label{101}
\mathcal E &=& \int 
\psi^{\dagger}(\beta m + \vec \alpha \cdot (\vec p + e \vec A)) \psi\ \de^3 x  
+ \int \frac{1}{2}(\vec B^2 + \vec E_T^2)\de^3 x 
\nonumber \\&&
- \frac{e^2}{4 \pi} 
\int \ \frac{\psi^{\dagger}(x) \psi(x) 
\psi^{\dagger}(x') \psi(x')}{|x-x'|}\ \de^3 x'\ \de^3 x.
\end{eqnarray}
Recalling that $\vec p + e \vec A $ is the kinetic momentum, we see 
now a clear separation into kinetic energy of the electron, 
kinetic energy of the photon (only $\vec E_T$ propagates) and 
an interaction part (in this case, self-interaction). 

In general, we can summarize our conclusions as follows. 
In the gravitational approach, the potential energy, and/or 
self-interaction, or longitudinal electric field contributions, are 
contained in the stress-energy tensor of the electromagnetic field, 
while the stress-energy tensor of the Dirac field only contains kinetic 
energy. 

On the other hand, using the canonical Noether stress-energy tensor, 
the same energy contributions  appear in the Dirac 
part of the stress-energy tensor, while the electromagnetic part 
contains only the propagating modes of the photon field.  

As a result, if one deals with electrons in background fields, i.e., 
if one omits the free Maxwell part of the Lagrangian, then one 
will have to use the canonical stress-energy tensor in order to 
derive a Hamiltonian  (because else, the interaction part will be missing). 

If, on the contrary, one deals with photons propagating on a certain 
background electron density, i.e., if one omits the free Dirac Lagrangian, 
one will still have to use the canonical stress-energy tensor (as we 
will show below), in order to find a conserved energy. Apart from 
the free Maxwell Lagrangian, one will have to add the interaction 
part $\bar \psi e \gamma^m A_m \psi$, which will lead to 
quite a different interaction contribution in the stress-energy tensor.

However, we still have to show that this prescription really leads 
to a conserved energy definition. This is not hard to do. Consider 
a Lagrangian  depending  on two fields $q,p$, where $p$ is considered 
to be a background field. Thus, the free field Lagrangian for the 
field $p$ is missing. (Imagine, e.g.,  $q = (\psi, \bar \psi)$ and $p = A_i$
 for the electron in a background electromagnetic field). 
Then, we have (flat limit) 
\begin{equation}\label{102}
\partial_i  L  = \frac{\partial L}{\partial q} q_{,i} + 
\frac{\partial L}{\partial q_{,k}} q_{,k,i} + 
\frac{\partial L}{\partial p} p_{,i}.  
\end{equation}
 where we suppose that the interaction part does not contain derivatives 
of the fields (therefore, no derivatives of $p$ are contained in $L$). 
We use the field equation for the field $q$, and find 
\begin{equation}\label{103}
0 = \partial_k [\tau^k_{\ i}] + \frac{\partial L}{\partial p}\ p_{,i}.
\end{equation}
For instance, for the fermion in the background electromagnetic field, 
we have $L = L_D$ and (\ref{103}) reads 
\begin{equation}\label{104}
0 = \partial_k [\tau^k_{D i}] + \frac{\partial L_D}{\partial A_m} A_{m,i}.
\end{equation}
The second term can alternatively be written in the form $- j^m A_{m,i}$. 
The integral form of the conservation law  reads 
\begin{equation}\label{105}
\frac{\de }{\de t} \int \tau_{\ i}^0 \ \de^3 x = \frac{\de \pi_i }{\de t} 
= - \int \frac{\partial
  L}{\partial p} p_{,i}\ \de^3 x,  
\end{equation}
or, in the Dirac case, 
\begin{equation}\label{106}
\frac{\de \pi_i }{\de t} =  \int j^m  A_{m,i}\ \de^3 x.   
\end{equation}
This, however, is exactly what is usually understood under \textit{canonical}
momentum: The component $\pi_i $ (for some fixed $i$) is conserved 
whenever the exterior field $p$ (or $A_m$) is independent of $x^i$. 
Especially, 
the energy $\pi_0$ is conserved whenever the exterior field is time 
independent. The same argument holds of course if one reverses the role 
of $A_i $ and $\psi$. 

On the other hand, with the gravitational stress-energy tensor, 
no conservation law can be formulated for the 
description of  a field on the background configuration of another field, 
since the derivation of the conservation law, based on coordinate 
invariance, relies on the complete matter equations of all 
fields (see section 2). This is rather disappointing, because the 
gravitational approach has the advantage that each part of the stress-energy 
tensor is by itself gauge invariance. As opposed to this, with a 
relation of the form (\ref{106}), one will have to make a suitable gauge 
choice in order to find reasonable results. (Already the statement, that 
$A_i$ is time independent will depend on the gauge one adopts.)

Finally, in order to avoid misunderstandings, we should say that 
there is a little bit more 
involved in the construction of the field Hamiltonian than just writing down 
an expression for the energy. Especially, we have to express the results 
in terms of the canonical variables (the fields and their conjugate momenta), 
which involves the use of a gauge fixing term in the Maxwell case. However, 
those manipulations are found in any textbook on quantum field theory and 
do not affect our specific arguments. 

Before we close this section, we wish to make a last remark concerning 
the positivity of our expressions for the energy. 
One reason for 
having confined the previous considerations to the flat limit was the 
fact that we then have $T^{00} = T_{00} = T^0_{\ 0}$ (where each index may 
be interpreted either as tangent or as spacetime index),  while in the 
general case, from a tensor $T^i_{\ a}$, there are quite a few possibilities
to define the energy density. 

For simplicity, we confine our discussion to general relativity and 
the metric stress-energy tensor $T^{ik}$ (which is equivalent to 
the canonical tensor, see (\ref{13})). 
It has been pointed out\cite{5} that 
$T_{00}$ is always positive (for reasonable matter Lagrangians), 
while $T^0_{\ 0}$, 
in general, has no definite sign. One is therefore tempted to consider 
$T_{00}$ to be the correct  energy density. We wish to point out that 
this looks like  the correct answer to the wrong question. 
If we suppose that $g_{ik}$ 
is diagonal, for simplicity, then $T^0_{\ 0}$ becomes  negative if 
$g_{00}$ is negative. This is the case, for instance, at the region inside  
the Schwarzschild horizon. 

Suppose we have some general conservation law $j^i_{,i} = 0$, where 
$j^i$ is not necessarily a tensor. Then, we can write
\begin{equation}\label{107}
0 = \int j^i_{,i}\ \de^3 x = \frac{\de }{\de t} \int j^0 \ \de^3 x + 
\int \vec \nabla \cdot \vec j \ \de^3 x,
\end{equation}
and by converting the last term into a surface term, we find that $\int j^0
\de^3 x $ is constant in time (i.e., in $t = x^0$). 

However, as we have pointed out in section 6, at the horizon, $t$ 
will become spacelike, and $r$ takes over the role of the time coordinate. 
Then, our conservation law does not seem to make much sense anymore!

Indeed, from $j^i_{,i}= 0 $, one should proceed as follows: 
\begin{equation}\label{108}
0 = \int j^i_{,i} \de^4 x = \oint j^i \de S_i,  
\end{equation}
where in the last expression, we have an integral over a  closed 
spacelike hypersurface. Only if $x^0$ is the time coordinate, 
we can conclude from this that $\int j^i \de S_i $ is independent 
of $x^0$ (see Ref. 
\cite{5}, \S 29), and that it is equal to $\int j^0 \de^3 x$. 
However, if the metric is such, that there is no clear separation between 
timelike and spacelike coordinates (sometimes, lightlike coordinates are 
used), of if simply, say,  $x^1$ is the timelike coordinate (as is 
the case inside the horizon), then those expressions have to be 
changed. In the second case, e.g., we can derive a conservation law in 
the form, say, 
$(\de / \de x^1)  \int j^1 \de x^0 \de x^2 \de x^3 = 0$. 

Our point is that, although $T^0_{\ 0}$ might be  negative   
inside the horizon, while $T_{00}$ is generally positive,  
this is not really an argument in favor of $T_{00}$ and against $T^0_{\ 0}$, 
because, as we saw in our specific example, it might well be some other 
component of $T^i_{\ k}$ that enters the energy conservation law and 
therefore, ultimately, the Hamiltonian.    

\section{Conclusions} 

The relation between the canonical Noether and the gravitational 
stress-energy tensors were investigated and the following results 
were established. 

In general relativity, if the metric derivatives do 
not couple directly to the matter fields (as is the case for 
the known bosonic matter in the standard model), both tensors are 
physically equivalent, in the sense that they lead to the same 
conservation law and the same momentum vector. In the unlikely case 
where the metric derivatives couple to matter fields, the equivalence 
holds only in the flat limit, i.e., if gravitational interactions are 
neglected. 

In the reformulation of gravity in terms of tetrad fields, which allows 
for the coupling of spinor matter, we have the same situation. For 
bosonic matter, the tensors are always equivalent, while for spinor 
fields, which couple to the tetrad derivatives via the spin connection,  
the equivalence is restored only in the flat limit. 

In Poincar\'e gauge theory, the results are similar. 
For bosonic matter fields, coupling only to the tetrad field,  
we have again equivalence between canonical and gravitational tensors, 
while for spinor fields, coupling directly to the Lorentz connection, 
the equivalence holds only in the  case of a vanishing connection.  
For the special class of teleparallel theories  (i.e., theories with 
zero curvature and gravity described exclusively by torsion), this will 
always be the case. In general theories (e.g., Einstein-Cartan), it holds 
again in  the limiting case where we neglect the gravitational 
interactions.   

We also briefly introduced a canonical stress-energy tensor for 
both gravity and matter fields, which contains the canonical 
matter tensor, as opposed to the well known Landau-Lifshitz tensor 
which is based on the Hilbert tensor as far as the matter contributions 
are concerned. Although this  tensor (which 
can be found in a problem section in the textbook of Landau 
and Lifshitz) is not symmetric, we find that it 
has some attractive features and renders the concept of gravitational 
energy less obscure than the original Landau-Lifshitz approach, 
revealing better the similarities to other fields. Moreover, it 
turns out that it allows 
for a straightforward generalization to Poincar\'e gauge theory. 

Further, we have derived the explicit expression of the 
Hilbert tensor  for a point charge in an electromagnetic field and 
pointed out problems related to  the change of the nature of the   
coordinates at the horizon of a black hole, which can change from 
timelike to spacelike and vice versa. Nevertheless, in the flat 
limit, the correct special relativistic expression for the 
energy is found from this tensor. 

Finally, we studied in detail the Dirac-Maxwell system in a flat 
background, showing explicitely 
the equivalence of the total stress-energy tensor
 in both canonical and gravitational approaches, focusing mainly on the 
time component in order to find expressions for the field energy, which 
is the starting point for the construction of the field Hamiltonian.  
It is found that in the canonical approach, the interaction part 
of the Hamiltonian is found in the contributions of the tensor 
that stems from the Dirac Lagrangian, while in the gravitational 
approach, the same terms are found to originate from the Maxwell Lagrangian. 
This makes clear that, when dealing only with parts of a system, 
i.e., when one considers certain fields as non-dynamical background 
fields, the equivalence between both approaches breaks down. It is 
then shown in general that the correct expressions for energy (Hamiltonian) 
and field momentum are derived,  in such a case,  from the canonical tensor, 
although the result will necessarily be gauge dependent.

\section*{Acknowledgments}

This work has been supported by EPEAEK II in the framework of ``PYTHAGORAS 
II - SUPPORT OF RESEARCH GROUPS IN UNIVERSITIES'' (funding: 75\% ESF - 25\% 
National Funds).  

\appendix

\section{Landau-Lifshitz tensor in Poincar\'e gauge theory}

We briefly investigate the question, whether it is possible to 
find a conservation law similar to (\ref{51}), but containing the 
gravitational tensor $T^i_{\ k}$ (from (\ref{33})) instead of the canonical 
tensor $\tau^i_{\ k}$ as far as the matter part is concerned. This 
is indeed possible in many ways, and just as in general relativity, 
we need additional criteria to make a reasonable choice. 

For simplicity, we consider first the case of Einstein-Cartan theory, 
with field equation $G^i_{\ k} = T^i_{\ k}$. Following Landau and Lifshitz, 
we write 
\begin{equation}\label{a1}
\partial_i \left[e(-G^i_{\ k} + r^i_{\ k} + T^i_{\ k})\right]  = 0, 
\end{equation}
which is trivially satisfied if $r^i_{\ k}$ is a relocalization term, 
i.e., if we have $(e\  r^i_{\ k}) = r^{li}_{\ \ k,l}$, with $r^{li}_{\ \ k} = 
r^{[li]}_{\ \ k}$. Then, we interpret 
$\tilde t^i_{\ k} = -G^i_k + r^i_{\ k}$ as stress-energy tensor of the 
gravitational field. In order to fix the relocalization term, Landau 
and Lifshitz choose the following two criteria: Firstly, $\tilde t^{ik}$ 
should be symmetric and secondly, it should not contain second and higher
 derivatives of the metric. (It turned out, in general relativity, 
that, in order to 
achieve both requirements, one has to replace $e = \sqrt{-g} $ by $e^2 = 
-g$.) 

We have already argued, at the end of section 4, that the first requirement 
does not make much sense in a general Poincar\'e gauge theory, since 
the matter part $T^{ik}$ is itself asymmetric. (This does not mean, however, 
that it is not possible, in principle, to symmetrize $\tilde t^{ik}$.) 
We therefore lift this requirement. 

As to the second criteria, we see that 
already the choice $\tilde t^i_{\ k} =- G^i_{\ k}$ does not contain 
higher derivatives of the independent fields $(e^a_i, \Gamma^{ab}_{\ \ i})$. 
Thus, no relocalization is needed, and we can directly interpret $- G^i_{\ k}$ 
as stress-energy of the gravitational field. 

In more general theories, the term $-G^i_{\ k}$ in (\ref{a1}) will have to be 
replaced by the corresponding expression 
$- \frac{1}{e}\ (\delta \mathcal L_{grav}/\delta e^a_i)  e^a_k $, 
(with $\mathcal L_{grav} = e L_{grav}$)
and may 
contain higher derivatives of the tetrad field (if $\mathcal L_{grav}$ 
contains 
terms quadratic in the torsion). Therefore, we proceed as follows: We start 
with the relation (\ref{51}), $\partial_i[e(t^i_{\ k} + \tau^i_{\ k})]=0$, and 
replace, by means of Eq. (\ref{49}), $\tau^i_{\ k}$ with 
\begin{equation}\label{a2}
\partial_i[e(\tau^i_{\ k})] = \partial_i[e(T^i_{\ k} + \sigma_{ab}^i
\Gamma^{ab}_{\ \  k})], 
\end{equation}
where $e  \sigma_{ab}^{\ \ i} = \delta \mathcal L / 
\delta \Gamma^{ab}_{\ \ i} = 
-   \delta \mathcal  L_{grav} / \delta \Gamma^{ab}_{\ \ i}
 = - \partial \mathcal L_{grav} / \partial \Gamma^{ab}_{\ \ i} + 
\partial_m ( \partial \mathcal L_{grav}/
\partial \Gamma^{ab}_{\ \ i,m})$. The conservation law then takes the form 
\begin{equation}\label{a3}
\partial_i\left[e(t^i_{\ k} + T^i_{\ k}) - \frac {\partial \mathcal L_{grav}}
{\partial  \Gamma^{ab}_{\ \ i}}\Gamma^{ab}_{\ \ k} + 
(\partial_m \frac{\partial \mathcal L_{grav}}
{\partial \Gamma^{ab}_{i,m}}) \Gamma^{ab}_k\right]
= 0.
\end{equation}
In the last term, we can omit a relocalization term $\partial_m 
(\frac {\partial \mathcal L_{grav}}{\partial \Gamma^{ab}_{\ \ i,m}} 
\Gamma^{ab}_{\ \ k})$ 
(in view of $\partial \mathcal L_{grav} / \partial \Gamma^{ab}_{\ \ i,m} = 
2 \partial \mathcal L_{grav} / \partial R^{ab}_{\ \ mi}$), and we find 
\begin{equation}\label{a4}
0 = \partial_i \left[ e(t^i_{\ k} 
- \frac{\partial L_{grav}}{\partial \Gamma^{ab}_{\ \ i}}
\Gamma^{ab}_{\ \ k} - \frac{\partial L_{grav}} {\partial \Gamma^{ab}_{\ \ i,m}}
\Gamma^{ab}_{\ \ k,m} 
+ T^i_{\ k})\right],
\end{equation}
or simply 
\begin{equation}\label{a5}
0 = \partial_i [ e(\tilde t^i_{\ k} + T^i_{\ k})], 
\end{equation}
where explicitely, we have 
\begin{equation} \label{a6}
\tilde t^i_{\ k} = \frac{\partial L_{grav}}{\partial e^a_{m,i}} e^a_{m,k} 
+ \frac{\partial L_{grav}}
{\partial \Gamma^{ab}_{\ \ m,i}} \Gamma^{ab}_{\ \ m,k}
- \frac{\partial L_{grav}}{\partial \Gamma^{ab}_{\ \ i}}
\Gamma^{ab}_{\ \ k} - \frac{\partial L_{grav}} {\partial \Gamma^{ab}_{\ \ i,m}}
\Gamma^{ab}_{\ \ k,m}- \delta^i_k L_{grav}. 
\end{equation}
It is not hard to check that this tensor does not contain higher derivatives 
of $e^a_i$ and $\Gamma^{ab}_{\ \ i}$ and that, in the special case of 
Einstein-Cartan theory, it reduces again to $-G^i_{\ k}$. 

What we have actually done in passing from the canonical relation 
(\ref{51}) to 
 the relation (\ref{a4}) is, apart from relocalization terms, shifting 
the term $\sigma_{ab}^{\ \ i} \Gamma^{ab}_{\ \ k}$ from the matter 
part of the stress-energy to the gravitational part. This is quite similar 
to what happened with the four fermion Coulomb term in section 7. 

In Einstein-Cartan theory, and possibly also in other cases, the 
total stress-energy $\tilde t^i_{\ k} + T^i_{\ k}$ is  zero throughout. 
Especially, outside of the matter distribution, there will be no 
gravitational energy density. This is rather disappointing, especially 
in view of potential applications concerning gravitational waves. 
However, in the absence of other criteria for the choice of $r^i_{\ k}$ 
in (\ref{a1}), the choice $r^i_{\ k} = 0 $ is as good as any other and  
attempts to introduce an additional relocalization term in order to 
find a non vanishing energy density would be ad hoc and arbitrary. 
The only natural way to get a stress-energy tensor different from 
zero seems to be the canonical approach of section 4. 
The canonical tensor (\ref{51}) 
has the  additional advantage that $e^a_i$ and $\Gamma^{ab}_{\ \ i}$ are 
treated in a symmetric way, which is not the case with (\ref{a5}).

\end{document}